\documentclass[%
 reprint,
superscriptaddress,
amsmath,amssymb,
prb,
]{revtex4-2}

\usepackage{graphicx} 
\usepackage{dcolumn} 
\usepackage{bm} 
\usepackage{hyperref} 
\usepackage[group-separator={,}]{siunitx} 
\usepackage{textcomp} 
\usepackage[english, capitalise]{cleveref} 
\usepackage{setspace} 
\usepackage{xcolor} 

\definecolor{noteorange}{RGB}{255, 128, 0}
\definecolor{hellerot}{RGB}{255, 128, 128}
\definecolor{rosa}{RGB}{255, 192, 203}
\definecolor{green}{rgb}{0.0, 0.5, 0.0}

\usepackage{subcaption} 

\begin{document}

\preprint{APS/123-QED}

\title{
A novel spectroscopic probe of ultrafast magnetization dynamics in the extreme ultraviolet spectral range 
}
\author{Johanna Richter}
\email{johanna.richter@mbi-berlin.de}
\author{Somnath Jana}
\author{Robert Behrends}
\affiliation{%
 Max-Born-Institut für Nichtlineare Optik und Kurzzeitspektroskopie, Max-Born-Straße 2A, 12489 Berlin, Germany
}%
\author{Carl S. Davies}
\affiliation{%
FELIX Laboratory, Radboud University, Toernooiveld 7, 6525 ED, Nijmegen, The Netherlands
}%
\author{Dieter W. Engel}
\author{Martin Hennecke}
\author{Daniel Schick}
\author{Clemens von Korff Schmising}
\email{korff@mbi-berlin.de}
\affiliation{%
 Max-Born-Institut für Nichtlineare Optik und Kurzzeitspektroskopie, Max-Born-Straße 2A, 12489 Berlin, Germany
}%
\author{Stefan Eisebitt}
\affiliation{%
 Max-Born-Institut für Nichtlineare Optik und Kurzzeitspektroskopie, Max-Born-Straße 2A, 12489 Berlin, Germany
}%
\affiliation{Technische Universität Berlin, Institut für Optik und Atomare Physik, 10623 Berlin, Germany}

\date{\today}

\begin{abstract}
The development of spectroscopic techniques in the extreme ultraviolet (XUV) spectral range has significantly advanced the understanding of ultrafast interactions in magnetic systems triggered by optical excitation. In this work, we introduce a previously missing geometry that facilitates the observation of the ultrafast magnetization dynamics of magnetic systems with an out-of-plane magnetization grown on XUV opaque substrates.
This novel approach to probe ultrafast magnetization dynamics combines the magneto-optical Kerr effect with the strong dependence of a sample's reflectance near its Brewster angle.
It therefore works with linearly polarized light and does not require any additional polarizing optics.
We provide a comprehensive analysis of the technique by presenting both simulations and experimental data as a function of the energy and the polarization of the XUV probe radiation as well as of the delay time after optical excitation.

\end{abstract}

\maketitle

\section{Introduction}

Ultrafast spectroscopy in the extreme ultraviolet (XUV) has become a valuable experimental technique for probing the microscopic processes of nonequilibrium magnetization.
 The spectroscopic observable arises from the off-diagonal elements in the dielectric tensor, which is strongly enhanced at atomic resonances, and hence allows for element-specific measurements of the magnetization. A number of different experimental geometries have been introduced in recent years, tailored for ultrafast pump-probe experiments both at free-electron laser (FEL) and laboratory-based high-harmonic generation (HHG) radiation sources.  
 Certainly, one of the most established experimental geometries is the transverse magneto-optical Kerr effect (T-MOKE) \cite{la2009ultrafast,la-o-vorakiat_ultrafast_2012}. In T-MOKE one measures the difference in reflectance for opposite magnetization directions oriented perpendicular to the plane of incidence of p-polarized radiation. T-MOKE experiments have substantially increased our microscopic understanding of spin-transfer processes between different sublattices or layers in metallic systems \cite{mathias_probing_2012,rudolf_ultrafast_2012,turgut_controlling_2013,tengdin_direct_2020, KorffSchmising2023, probst_unraveling_2024,Gupta2023}.
 T-MOKE has been extended by angle- \cite{turgut_stoner_2016,probst_unraveling_2024} and polarization-resolved \cite{zusin_direct_2018} measurements, allowing a reconstruction of the resonant dielectric tensor with off-diagonal components in non-equilibrium.
 In addition, by analyzing the magnetization-dependent reflectance as a function of either photon energy or angle of incidence, T-MOKE provides direct insight into the transient magnetization depth profiles \cite{Hennecke2022,Chardonnet2021}.
 At the same time, we note that the high information content of the T-MOKE observable inherently complicates its interpretation, and recent studies have shown that great care must be taken when relating observables to both static \cite{Richter2024} and transient magnetization \cite{jana_analysis_2020,probst_unraveling_2024}. 
 
Another class of spectroscopic techniques in the XUV spectral range is based on polarization analysis in conjunction with the magneto-optical Faraday or Kerr effect, which causes changes in the polarization state of the light when it interacts with the magnetic material. These techniques also works with linearly polarized radiation but rely on polarization analysis, which in the XUV spectral range adds significant complexity to the experiment. It has been realized in reflection and transmission geometries at both HHG \cite{Alves2019} and FEL sources \cite{yamamoto_ultrafast_2015,Yamamoto2020,KorffSchmising2020}, including the implementation of novel polarization concepts specifically designed for ultrafast measurements \cite{caretta_novel_2021,pancaldi_comix_2022}. 

Experiments using magnetic circular dichroism (MCD) measure magnetization through the absorptive part of the dichroic index of refraction, making it easier to relate it to the magnetization. Previous studies have investigated in-plane and out-of-plane magnetic systems, but have exclusively used a transmission geometry with broadband HHG radiation \cite{willems_probing_2015,Siegrist2019,willems_optical_2020,Yao2020_a, Geneaux2024}. Another very promising implementation of the MCD geometry is based on time-streaking via zone-plate optics, and has been pioneered at free-electron laser facilities \cite{jal_single-shot_2019,hennes_time-resolved_2020,rosner_simultaneous_2020}.
However, MCD measurements require circularly polarized radiation. This is an additional experimental challenge, despite the increasing availability of polarization control both in FEL sources by special undulators and in HHGs either by two-color fields driving the harmonic process \cite{Kfir2014,Fan2015, Lambert2015} or by special optical elements \cite{Vodungbo2011, KorffSchmising2017}. 

What is missing in magnetization-sensitive XUV spectroscopy is a simple, intensity-based technique to study the magnetization dynamics of samples with an out-of-plane anisotropy, which are grown on XUV opaque substrates. 
This includes, for example, epitaxial systems or any system where the growth process requires high temperatures incompatible with nanometer thin membranes as growth substrates, such as 2D magnets \cite{lopes_large-area_2021}. 
In addition, for high repetition rate XUV experiments, thermal accumulation can become a concern if the sample is highly absorbing, so that efficient heat dissipation may only be possible through the use of engineered substrates.
While magneto-optical effects satisfying these experimental requirements have been discussed in the past for soft x-rays \cite{oppeneer_2003}, the very low reflectivities in this spectral range have prevented any significant applications.
We address this shortcoming in the current work and present a new geometry to use in the XUV spectral range, P-MOKE, that yields large magnetic asymmetries for out-of-plane magnetic systems.  It requires only linearly polarized light incident at angles around \SI{45}{\degree} and does not rely on additional polarizing optics. Directly related, we present a second new geometry, L-MOKE, where the magnetization is oriented longitudinally, i.e. in the plane of the sample and in the plane of reflection.
Qualitatively, both effects are based on changes in the polarization state of the probing XUV light as it interacts with the magnetic film, combined with the strong polarization dependence of the reflectance near the Brewster angle. In other words, the sample itself acts as both polarizing and analyzing the XUV radiation.

\begin{figure*}[ht]
    \includegraphics[width=1\linewidth]{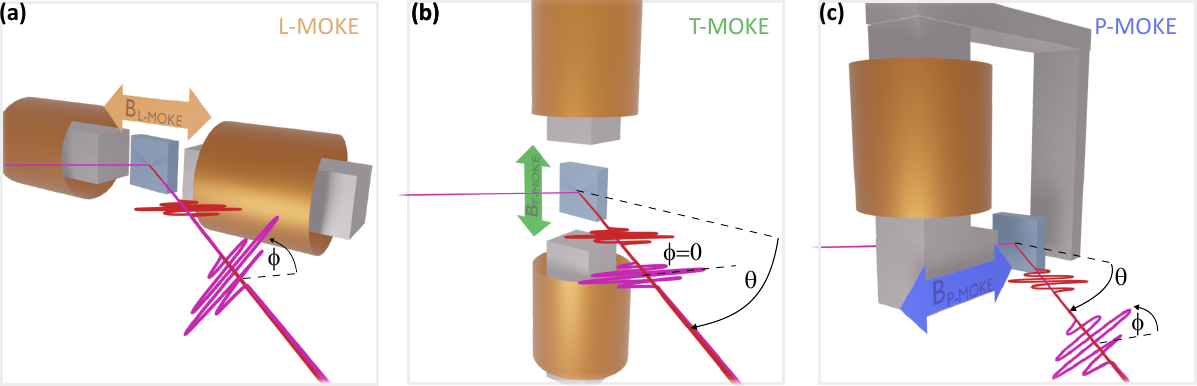}
    \caption{Illustration of the experimental geometries for the (a) longitudinal (L-MOKE), (b) transverse (T-MOKE),  and (c) polar (P-MOKE) magneto-optical Kerr effect. The respective alignment of the direction of the magnetization is controlled by an external magnetic field. For all three geometries the incidence angle is $\theta=\SI{45}{\degree}$. For L/P-MOKE the polarization angle $\phi \neq 0$, for T-MOKE we measure at  $\phi \approx 0$, i.e. for p-polarized radiation.  }
    \label{fig:main}
\end{figure*}

\section{Experimental details}
We study two sample structures: \\SiO$_{2}$/Ta(\SI{2}{nm})/Fe$_{50}$Ni$_{50}$(\SI{8}{nm})/Ru(\SI{3}{nm}) and \\SiO$_{2}$/Pt(\SI{3}{nm})/Gd$_{22}$Co$_{78}$(\SI{6}{nm})/Pt(\SI{3}{nm}).  Both samples were grown by magnetron sputtering. The FeNi film exhibits an in-plane magnetic anisotropy with a coercive field of 2.5 mT, whereas the Pt/GdCo/Pt sample exhibits an out-of-plane magnetic anisotropy with a coercive field of around 15 mT. 

Static and time-resolved measurements are performed using HHG radiation in the spectral range between 45 and \SI{72}{eV}. 
To produce such light, we focus near-infrared (NIR) laser pulses at a central wavelength of \SI{800}{nm}, a pulse duration of \SI{25}{fs} and a pulse energy of approximately \SI{2.5}{mJ} with a repetition rate of \SI{3}{\kilo\hertz} into a gas cell filled with neon. 
The HHG radiation produced is dominated by narrow-band emission peaks that are about \SI{3.1}{eV} apart and have a full width at half maximum of approximately \SI{0.2}{\electronvolt}. The peak energy values are determined with an uncertainty of $ \pm\SI{0.1}{\electronvolt}$. Quasi-continuous spectra are achieved by increasing the laser intensity beyond the ionization threshold of neon and varying the chirp of the laser pulse.
In order to set the polarization angle $\phi$ of the XUV radiation, i.e., to continuously vary it between linear p- and s-polarization, we rotate a $\lambda/2$-wave plate placed in the NIR laser beam in front of the HHG cell.  After reflection off the sample the XUV light is spectrally dispersed by a flat field grating and detected on an in-vacuum CCD camera. For all explored geometries the incidence angle, $\theta$, is set to \SI{45}{\degree}.
We define our experimental observable, the magnetic asymmetry, as the normalized difference of the reflectance, $R^{\uparrow,\downarrow}$, for two opposite magnetization directions set by an electromagnet: 
\begin{equation}\label{eq:A}
    A = \frac{R^{\uparrow} - R^{\downarrow}}{R^{\uparrow} + R^{\downarrow}} 
\end{equation}
To improve the signal-to-noise-ratio, we normalize the measured intensities with a second spectrometer placed in front of the sample and toggle the magnetization direction with a frequency of $\SI{1}{Hz}$.
The time-resolved data are obtained by measuring the asymmetry for different time delays after optical excitation with fluences of $\approx \SI{15}{mJ/cm\squared}$ for the FeNi sample and $\approx\SI{20}{mJ/cm\squared}$ for Pt/GdCo/Pt sample. The pump pulses have  at a center wavelength of \SI{800}{nm} and a pulse duration of $\approx\SI{50}{fs}$. 
Further details of the experimental setup are summarized in Ref. \cite{yao_tabletop_2020}.

\section{Spectra in L/T/P-MOKE}
Schematic illustrations of the three different experimental geometries are shown in Fig.~\ref{fig:main}. 
The geometries are distinguished by two experimental parameters: first, the orientation of the magnetization with respect to the plane of reflection of the probing XUV radiation and second the polarization angle $\phi$ of the XUV radiation. 
Both L- and T-MOKE are used to probe samples with an in-plane magnetization.
In the case of L-MOKE, the magnetization is set to lie within the plane of reflection, while in T-MOKE, the magnetization is perpendicular to plane of reflection. 
P-MOKE probes the out-of-plane magnetization, which again lies within the plane of reflection. 
For T-MOKE the probing XUV radiation is set to p-polarization ($\phi$=0), while for L-MOKE and P-MOKE the plane of polarization is rotated by a small angle $\phi$.  
\begin{figure}
    \includegraphics[width=\linewidth]{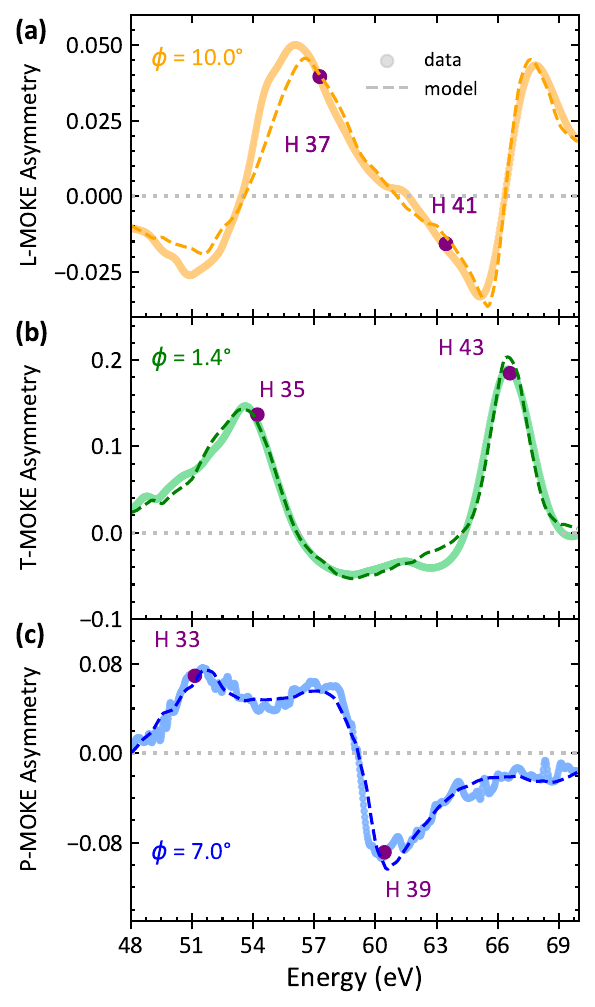}
    \captionsetup{justification=raggedright} 
    \caption{Measured and modeled asymmetry spectra in (a) L-MOKE and (b) T-MOKE geometries for the FeNi sample and in (c) P-MOKE geometry for the Pt/GdCo/Pt sample. The spectra were acquired at polarization angles, $\phi$, of a) \SI{10.0}{\degree}, (b) \SI{1.4}{\degree} and (c) \SI{7.0}{\degree} and at an angle of incidence $\theta=\SI{45}{\degree}$. The dots mark the energies of the odd HHG orders, H33 to H43, for which we have measured polarization- and time-dependent data.
    }
    \label{fig:spectra}
\end{figure}
    
In \cref{fig:spectra} the asymmetry spectra of FeNi and Pt/GdCo/Pt are shown as a function of photon energy. Two different geometries for FeNi are shown: (a) L-MOKE and (b) T-MOKE. Both spectra were measured on the same sample, but with the corresponding longitudinal and transverse magnetization orientation and for different polarization angles of $\phi=\SI{10.0}{\degree}$ (L-MOKE) and $\phi = \SI{1.4}{\degree} \approx \SI{0}{\degree}$ (T-MOKE).  The dashed lines are magnetic reflectance simulations based on a matrix formalism implemented in the \textsc{udkm1Dsim} Python package \cite{schick_udkm1dsim_2021,Elzo2012}. 
This toolbox allows us to define the magnetic properties of the sample structure and probe geometry and, with a recent extension, to specify the angle of polarization of the incident radiation.
The reflectances and hence the asymmetries depend sensitively on structural parameters such as thickness, roughness, density, stochiometry, as well as on the material-dependent atomic and magnetic form factors of each material layer. 
To minimize the residuals between experimental data and simulations, we allow for variations in densities as well as absolute values of the magnetization amplitude but keep the thicknesses and stochiometries fixed. All interface roughnesses are set to zero. The atomic form factors are taken from \citet{Henke1993}, while the magnetic form factors are taken from \citet{Willems2019}.  
We find a good agreement between experiment and simulation for both geometries and can clearly discern the M$_{2,3}$-edge for Fe around \SI{52.7}{eV} and for Ni around the energies of the $M$-edges M$_3$=\SI{66.2}{\electronvolt}, M$_2$=\SI{68}{\electronvolt}
\cite{booklet2001x}. Note that the asymmetry also reaches significant values of up to 5\% in L-MOKE. 

Fig.~\ref{fig:spectra}(c) displays the measured and modeled asymmetries of the Pt/GdCo/Pt sample in P-MOKE geometry. Again, the simulation is based on the \textsc{udkm1Dsim} toolbox and we vary the same set of parameters as mentioned above. Large asymmetry values exceeding 8\% are found both at the Co M$_{2,3}$ resonance around \SI{60}{eV} and at the Pt O$_{3}$ resonance around \SI{51.7}{eV} \cite{KorffSchmising2023}. We emphasize that in the case of P-MOKE the magnetic asymmetry is measured for an out-of-plane magnetization using linearly polarized light without an additional analyzer.

The solid dots in Fig.~\ref{fig:spectra} mark the photon energies of the odd HHG emission peaks H33 to H43, for which we will show polarization- and time-dependent data later.

\section{Polarization-dependent measurements}
    \begin{figure}[ht]
        \includegraphics[width=\linewidth]{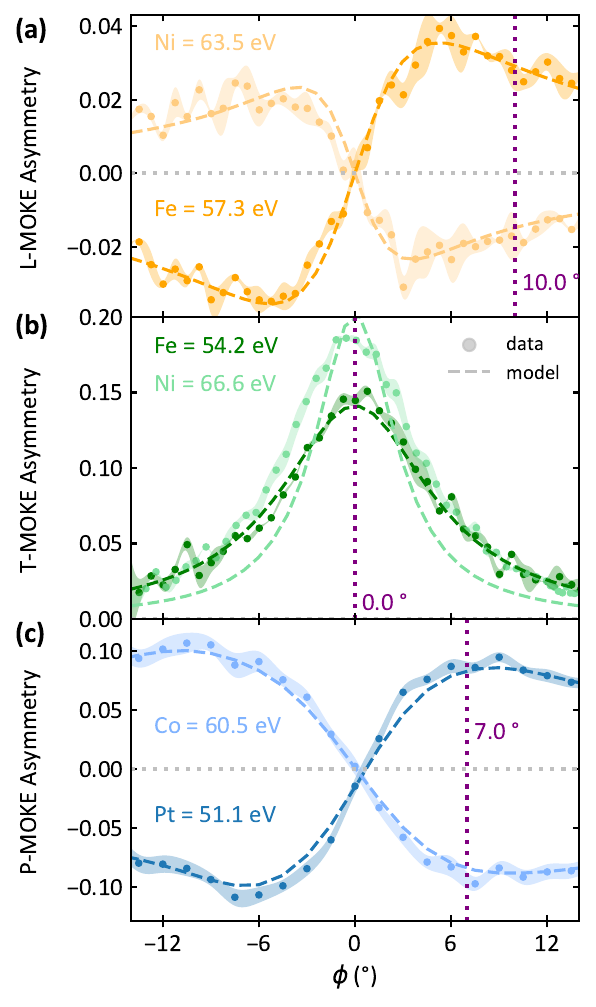}
        \captionsetup{justification=raggedright} 
        \caption{Measured and simulated magnetic asymmetries as a function of polarization angle $\phi$ for FeNi in (a) L-MOKE and (b) T-MOKE geometry, as well as for (c) Pt/GdCo/Pt in P-MOKE geometry. The data for the FeNi sample is shown for photon energies H37 = \SI{57.3}{\electronvolt} and H42 = \SI{65.3}{\electronvolt} as well as H35 = \SI{54.2}{\electronvolt} and H43 = \SI{66.6}{\electronvolt} with a predominant sensitivity to Fe and Ni, respectively.  For Pt/GdCo/Pt, we show data at H33 = \SI{51.1}{\electronvolt} and H39 = \SI{60.5}{\electronvolt}, which probe the magnetization of Co and Pt, respectively. The shaded bands represent the standard error of the measurement.}
        \label{fig:pol}
    \end{figure}
In the following, we examine in more detail, how the asymmetries in the three different geometries depend on the rotation angle $\phi$ of the plane of linear polarization of the probe radiation, recalling that $\phi=\SI{0}{\degree}$ corresponds to p-polarization and $\phi = \SI{90}{\degree}$ corresponds to s-polarization.  
The asymmetry as a function of $\phi$ are collected in Fig.~\ref{fig:pol} for L-, T- and P-MOKE.  
The solid dots are measured data, the dashed lines are calculated values using the identical structural and magnetic parameters as in the asymmetry spectra shown in Fig.~\ref{fig:spectra}) without any further adjustments.
The dotted vertical line represents the polarization angle for which we measured the time-resolved data. 
For the L-MOKE geometry, shown in panel (a), we show two exemplary measurements at the HHG emission peaks H37 and H41, corresponding to photon energies sensitive to Fe and Ni, respectively. 
For both energies, the asymmetry has a zero crossing for exact p-polarization and reaches its maximum absolute magnitude at about $\pm\SI{5}{\degree}$. 
For larger angles $\phi$ the asymmetries slowly approach zero again. 
We find a good agreement between experiment and model.

For the T-MOKE geometry, shown in \cref{fig:pol}(b), we find a qualitatively different behavior, well known from previous studies: the asymmetries are maximum for exact p-polarization and drop towards zero for a rotated polarization $\phi \neq 0$.   
The simulations match the intensity maxima; however, the modeled width at the Ni resonance is narrower than in the experimental data.

The P-MOKE asymmetry (see \cref{fig:pol}(c) as a function of the polarization angle follows the same qualitative behavior as for the L-MOKE case, with a zero crossing of $A$ at $\phi = 0$ and a maximum/minimum signal around $\phi \approx \SI{8}{\degree}$. This strongly suggests that both L-MOKE and P-MOKE asymmetries have the same physical origin. 

\section{Origin of P/L-MOKE asymmetry }
\begin{figure*}[ht]
    \includegraphics[width=\linewidth]{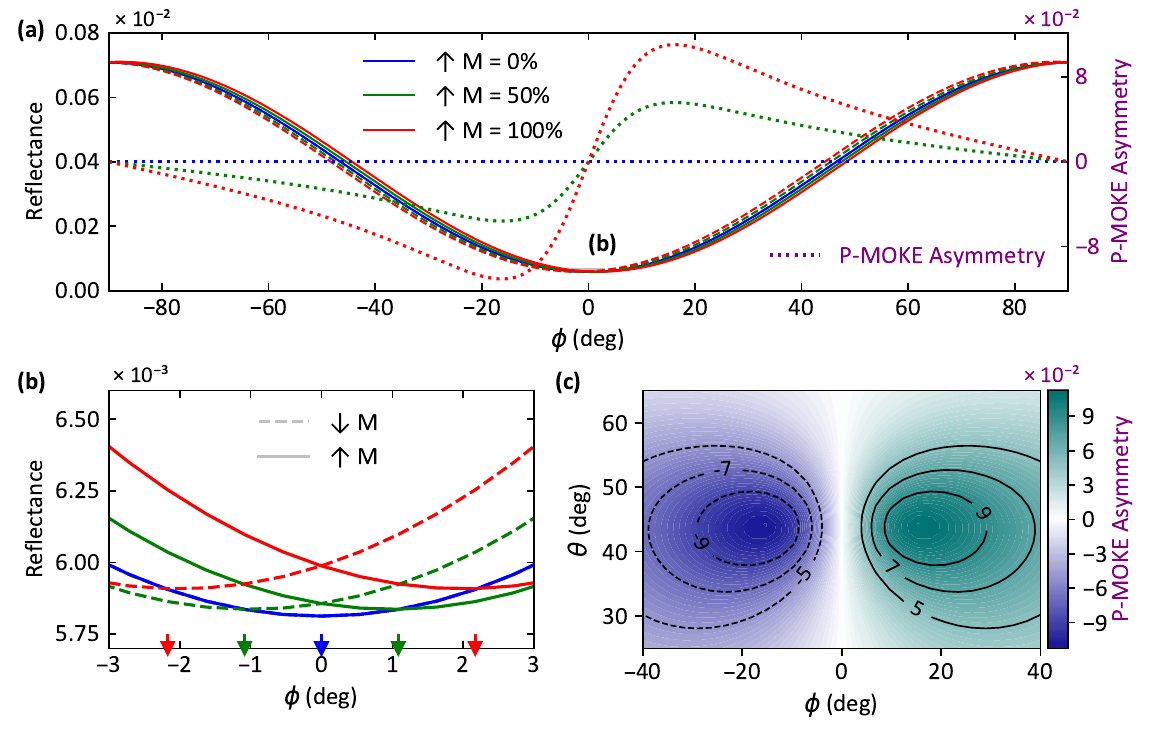}
    \captionsetup{justification=raggedright} 
    \caption{(a) Simulations of the reflectance and asymmetry as a function of the polarization angle $\phi$ of the Pt/GdCo/Pt sample in the P-MOKE geometry at a photon energy \SI{60.5}{\electronvolt}.  The calculation is shown for three different values of the magnetization, $M=0\%, 50\%$ and 100\%. Panel (b) focuses on $\phi$ angles close to zero with arrows indicating the location of the minimum reflectance for the different magnetization values. Panel (c) shows the P-MOKE asymmetry as a function of both  the incident angle $\theta$ and the polarization angle $\phi$.}
     \label{fig:PMOKE}
\end{figure*}

\begin{figure}[ht]
    \includegraphics[width=\linewidth]{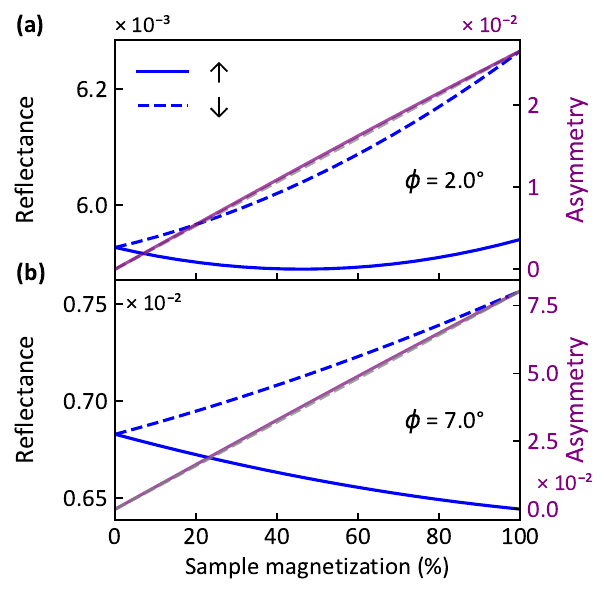}
    \captionsetup{justification=raggedright} 
    \caption{Simulations of the reflectance and asymmetry for the Pt/GdCo/Pt sample as a function of varying magnetization in P-MOKE geometry at $\theta = \SI{45}{\degree}$ for polarization angles (a) $\phi = \SI{2.0}{\degree}$ and (b) $\phi = \SI{7.0}{\degree}$. }
    \label{fig:mag}
\end{figure}
In this section, we perform further simulations to develop a better understanding of the origin of the magnetization sensitivity in XUV L-MOKE and P-MOKE.
In Fig.~\ref{fig:PMOKE}, we show the calculated reflectance, $R^{\uparrow,\downarrow}$, of the Pt/GdCo/Pt sample at a photon energy of $\SI{60.5}{eV}$ at an incidence angle of $\theta = \SI{45}{\degree}$ as a function of the polarization angle $\phi$. We calculate values for two opposite magnetization directions (solid and dashed lines) and for different amplitudes of the magnetization: blue for 0\% magnetization, green for 50\% magnetization, and red for 100\% magnetization. We stress that the calculation is performed for an equilibrium demagnetization that is spatially homogeneous along the depth of the sample. The absolute value of the reflectance increases from 0.3\% to 4\% as the light becomes s-polarized. 
In the same panel, we also plot the corresponding asymmetry, $A$ (dotted lines). The asymmetry increases as we rotate the polarization away from $\phi$=0, reaches a maximum at about $\phi = \pm\SI{10}{\degree}$ and finally approaches zero again for s-polarization at $\phi = \pm\SI{90}{\degree}$.
We note that even though there is a region of maximum asymmetry, for radiation sources with a limited flux, it may be advantageous to measure at larger angles, $\phi$, where the reflectance is larger. 
In panel (b), we focus on a region around $\phi=\SI{0}{\degree}$, where the polarization is close to p-polarization. 
The minima of the reflectance curves for the two magnetization directions are shifted by a characteristic Kerr angle $\phi_\mathrm{K} = \pm \SI{2.16}{\degree}$, as highlighted by the red arrows in Fig.\ref{fig:PMOKE}(b). 
As the magnetization is decreased to 50\% , $\phi_\mathrm{K}$ reduces to $\pm\SI{1.08}{\degree}$ (green arrows in Fig.\ref{fig:PMOKE}(b)) and finally for $M = 0$ the reflectance curves overlap  with their minima at $\phi = \SI{0}{\degree}$. Evidently, the angle $\phi_\mathrm{K}$ is proportional to the magnetization. 
Since the normalized difference in reflectance represents the asymmetry, as introduced in \cref{eq:A}, we can conclude that the asymmetry for P-MOKE is zero when the incident radiation is p-polarized ($\phi=\SI{0}{\degree}$), because the reflectance curves for opposite magnetization directions intersect at this point.
The panel \ref{fig:PMOKE} (c) shows the P-MOKE asymmetry of the Pt/GdCo/Pt sample in relation to the incident angle $\theta$ and the polarization angle $\phi$. One can appreciate that the asymmetry is antisymmetric around $\phi = \SI{0}{\degree}$ with a maximum at an incident angle $\theta=\SI{45}{\degree}$. It is also noteworthy that in a broad angle range of $\theta = \SIrange[]{40}{50}{\degree}$ and $\phi = \SIrange[]{10}{30}{\degree}$ the P-MOKE asymmetry remains $>8\%$. As the Brewster angle in the XUV spectral range varies only slightly with photon energy and material, the dependence of the incident angle $\theta$ in Fig.~\ref{fig:PMOKE} (c) reflects a general P-MOKE characteristic.

The observed behavior of the reflectance, with its minimum shifting as a function of the polarization angle, resembles the response of the optical Kerr effect, as measured by polarizing optics \cite{steinbach_wide-field_2021}. Here, the Kerr rotation of the optical radiation upon reflection off the magnetic sample is determined by measuring the transmission through a polarizer as a function of its rotational alignment.
By analogy, in P-MOKE, the nonzero projection of the XUV radiation \textbf{k}-vector onto the magnetization vector induces a finite Kerr rotation. This rotation is analyzed by reflecting at the Brewster angle, leading to a magnetization-dependent reflectance.

In \cref{fig:mag}, we illustrate the behavior of the reflectance and asymmetry as a function of the amplitude of the magnetization for two fixed angles $\phi$.  
In panel (a), we show the scenario for $\phi = \SI{2}{\degree}$: evidently the reflectance varies in a nonlinear fashion as the amplitude of the magnetization is reduced. 
For the case of the magnetization pointing up, the reflectance even exhibits a non-monotonic behavior, it first decreases, reaches a minimum for a magnetization amplitude of about 50\% before it increases again. 
For larger values of $\phi$, the reflectance varies approximately linearly as the amplitude of the magnetization is changed, cf. panel (b). 
We would like to emphasize once again that this is the same qualitative response as observed in optical measurements. 
Particularly in time-resolved Kerr microscopy, where a camera detects intensities after polarization analysis, it is often overlooked that following the reflectance for only one magnetization direction is insufficient to infer the transient magnetization state \cite{hashimoto_ultrafast_2014,steinbach_wide-field_2021}. Importantly, the asymmetry (indicated in purple in Fig.\ref{fig:mag}) itself is for this sample structure almost linear as a function of magnetization, independent of the polarization angle $\phi$.  We find a very small relative deviation from a perfect linear relationship between asymmetry and magnetization (dashed grey line) with a maximum of 2\% at 50\% magnetization at the E=\SI{60.5}{\electronvolt} and  $\phi = \SI{7}{\degree}$. 

Furthermore, we have computationally studied the relationship between the magnetic asymmetry and an equilibrium as well as spatially homogeneous demagnetization for different photon energies, polarizations, and angles of incidence, and for various typical sample structures. 
We find that for magnetic films capped with a heavy metal such as tantalum, platinum, or ruthenium with a thickness of at least \SI{2}{nm}, the asymmetry has a linear dependence on the magnetization amplitude.  
In bare magnetic films, or films coated with a light element or a native oxide layer, we simulate P-MOKE asymmetries exceeding 80\% for specific values of the parameters $E$, $\phi$, and $\theta$. In these cases a linear relationship between magnetization and asymmetry is not always satisfied.
Such nonlinearities arise when changes of the reflectance upon demagnetization are not dominated by a shift of $R^{\uparrow \downarrow}(\phi)$ by an decreasing angle $\phi_\mathrm{K}$, but rather by absolute changes of reflectance, i.e. the minima of $R^{\uparrow \downarrow}(\phi)$ remain constant at $\phi \approx \SI{90}{\degree}$. 
We emphasize that this limitation also applies to the T-MOKE geometry \cite{Richter2024}. 

Our simulations also suggest a strategy to increase the P/L-MOKE asymmetries while maintaining a linear relationship between asymmetry and magnetization. Tuning the thickness of the magnetic film to $d\approx \lambda \sin(\theta)/2$, where $\lambda$ is the resonant XUV wavelength, maximizes the asymmetry. We explain this by constructive interference between light reflected from the back and front side of the magnetic film. Generally, we find that the shape of the asymmetry spectrum is strongly influenced by the sample's geometry, with its maxima moving in the parameter space defined by the photon energy, $E$, polarization angle $\phi$ as well as incidence angle, $\theta$.

\section{Ultrafast magnetic response in L/T/P-MOKE}

    \begin{figure}[ht]
    \includegraphics[width=\linewidth]{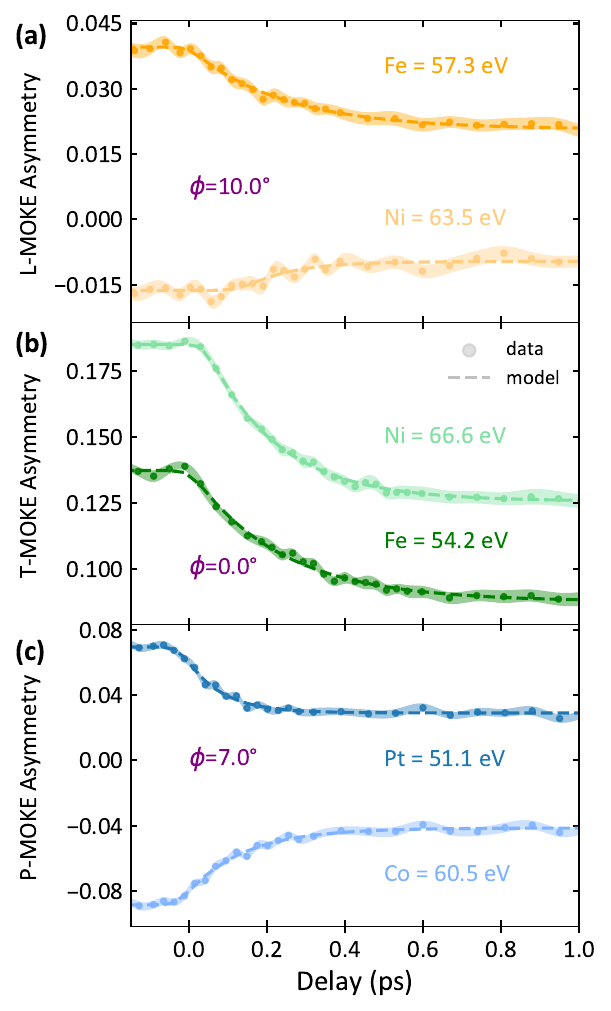}
    \captionsetup{justification=raggedright} 
    \caption{Time-resolved magnetization dynamics measured in (a) L-MOKE and (b) T-MOKE geometries for FeNi, and (c) in P-MOKE geometry for Pt/GdCo/Pt. 
    The shaded bands represent the standard error of the measurement.}
    
    \label{fig:delay}
    \end{figure}
    
In this section, we present time-resolved measurements for all three discussed geometries.
In \cref{fig:delay}(a) and (b), we show the ultrafast dynamics of the FeNi alloy for both L-MOKE and T-MOKE. For both cases, we select the same two photon energies as in \cref{fig:pol}, representing the response of Fe and Ni at the indicated $\phi$ angle. The data are fitted with a single exponential decay function describing the ultrafast reduction of the asymmetry, with the time constant $\tau$ amplitude $C$ and static asymmetry $A_0$:

\begin{equation}
    A(t) =  A_0 + G(t) \ast \left\{ \Theta(t-t_\mathrm{0}) C \left(1-e^{-(t-t_\mathrm{0})/\tau} \right) \right\}
    \label{eq:exp} 
\end{equation}

We consider the temporal resolution in our experiment by a convolution with a Gaussian function, $G(t)$, with a full width at half maximum of \SI{50}{fs}, corresponding to the cross-correlation of the XUV and excitation pulse durations. $\Theta$ is the Heaviside function and $t_\mathrm{0}$ defines the onset of the magnetization dynamics.
We find decay times for Fe/Ni of $\tau=(\SI{242}{fs}\pm\SI{19}{fs})/(\SI{119}{fs}\pm\SI{60}{fs})$ and $\tau=(\SI{236}{fs}\pm\SI{14}{fs})/(\SI{160}{fs}\pm\SI{9}{fs})$ for L-MOKE and T-MOKE, respectively. Closer inspection also reveals a delay ($t_0^\mathrm{Ni}-t_0^\mathrm{Fe}$) of $(\SI{128}{fs}\pm\SI{39}{fs})$ and $(\SI{38}{fs}\pm\SI{8}{fs})$ between the onset of the Fe and Ni magnetization for the two respective geometries, L- and T-MOKE. The errors correspond to a 68\% confidence interval of the fits.

Both geometries reproduce the trend of a faster demagnetization of Ni compared to Fe as well as confirm the delayed onset of the Ni demagnetization. While the microscopic origin of this delay is still disputed in literature, the observation itself has been confirmed in many independent measurements \cite{mathias_probing_2012,jana_setup_2017,jana_experimental_2022,hofherr_ultrafast_2020,moller_verification_2024,korffschmising_2024}. 
The slight differences between the extracted fitting parameters in the L- and T-MOKE geometries deserve a brief discussion. First, the L- and T-MOKE measurements were performed in two separate experiments, with slightly different excitation conditions. 
Secondly, we probe the Fe and Ni response at different photon energies. This is relevant, since it has been shown before that for non-equilibrium magnetization dynamics, different probing photon energies can lead to different response functions \cite{jana_analysis_2020, Yao2020_a, hennes_time-resolved_2020}. 
In addition, some of us have recently shown that optically excited magnetic heterostructures can form a transient magnetization depth profile leading to shifted and reshaped asymmetry spectra \cite{Hennecke2022}. The different shapes of the asymmetry spectra for L- and T-MOKE (cf. \cref{fig:spectra}) can therefore lead to energy-dependent changes of the asymmetry.  

In \cref{fig:delay}(c), we display the dynamics of Co and Pt in the Pt/GdCo/Pt sample. 
We again describe the data with Eq.~\ref{eq:exp}, and extract demagnetization time constants of $\tau=\SI{145}{fs}\pm\SI{8}{fs}$ and $\SI{89}{fs}\pm\SI{7}{fs}$ for Co and Pt, respectively. 
As an aside, we note that the distinct ultrafast response of intrinsic and induced moments has puzzled researchers for over a decade, with many conflicting results published in the literature \cite{kuiper_spin-orbit_2014,hofherr_induced_2018,willems_optical_2020,yamamoto_ultrafast_2019,Yamamoto2020,vaskivskyi_element-specific_2021,hennes_element-selective_2022,KorffSchmising2023}. While it is beyond the scope of the present work to elaborate on this interesting physics, we would like to mention that the observation of a faster demagnetization dynamics of Pt compared to Co is in agreement with our previous work based on MCD \cite{willems_optical_2020} and T-MOKE \cite{KorffSchmising2023}.

In conclusion, we introduced two new geometries for ultrafast studies of magnetization using XUV radiation generated by an HHG-based light source, L-MOKE and P-MOKE. We combined experimental and computational data to characterize the respective observables as a function of both photon energy and polarization angle. 
We show that the magnetic asymmetry of L- and P-MOKE scales linearly with the amplitude of the magnetization, making it suitable for ultrafast studies. 
A qualitative explanation based on Kerr rotation and successive polarization-dependent reflections near the Brewster angle provides an intuitive understanding of our observations. 
In particular, we believe that P-MOKE -- with relatively large asymmetries on the order of 10\% -- will emerge as an useful new geometry for studying the ultrafast dynamics of magnetic systems with out-of-plane anisotropy that cannot be grown on XUV-transparent membranes. 
Furthermore, we anticipate that P-MOKE in the XUV spectral range will also find applications in static and time-resolved imaging of nanoscale magnetic domains and textures.

\section*{Data availability statement}
All of the data presented in this article is publicly available \cite{dataJR}. Raw data and further details on the analysis are available upon reasonable request from the authors.

\begin{acknowledgments}
C. v. K. S., J.R. and S. E. acknowledge financial support from the Deutsche Forschungsgemeinschaft (DFG, German Research
Foundation) – Project-ID 328545488 – TRR 227, project A02. 
D.S. acknowledges funding by the Leibniz Association through the Leibniz Junior
Research Group Grant No. J134/2022.
This project has received funding from the European Union's Horizon 2020 research and innovation programme under grant  agreement no. 871124 Laserlab-Europe. 
\end{acknowledgments}

\bibliography{references}

\begin{thebibliography}{55}%
\makeatletter
\providecommand \@ifxundefined [1]{%
 \@ifx{#1\undefined}
}%
\providecommand \@ifnum [1]{%
 \ifnum #1\expandafter \@firstoftwo
 \else \expandafter \@secondoftwo
 \fi
}%
\providecommand \@ifx [1]{%
 \ifx #1\expandafter \@firstoftwo
 \else \expandafter \@secondoftwo
 \fi
}%
\providecommand \natexlab [1]{#1}%
\providecommand \enquote  [1]{``#1''}%
\providecommand \bibnamefont  [1]{#1}%
\providecommand \bibfnamefont [1]{#1}%
\providecommand \citenamefont [1]{#1}%
\providecommand \href@noop [0]{\@secondoftwo}%
\providecommand \href [0]{\begingroup \@sanitize@url \@href}%
\providecommand \@href[1]{\@@startlink{#1}\@@href}%
\providecommand \@@href[1]{\endgroup#1\@@endlink}%
\providecommand \@sanitize@url [0]{\catcode `\\12\catcode `\$12\catcode `\&12\catcode `\#12\catcode `\^12\catcode `\_12\catcode `\%12\relax}%
\providecommand \@@startlink[1]{}%
\providecommand \@@endlink[0]{}%
\providecommand \url  [0]{\begingroup\@sanitize@url \@url }%
\providecommand \@url [1]{\endgroup\@href {#1}{\urlprefix }}%
\providecommand \urlprefix  [0]{URL }%
\providecommand \Eprint [0]{\href }%
\providecommand \doibase [0]{https://doi.org/}%
\providecommand \selectlanguage [0]{\@gobble}%
\providecommand \bibinfo  [0]{\@secondoftwo}%
\providecommand \bibfield  [0]{\@secondoftwo}%
\providecommand \translation [1]{[#1]}%
\providecommand \BibitemOpen [0]{}%
\providecommand \bibitemStop [0]{}%
\providecommand \bibitemNoStop [0]{.\EOS\space}%
\providecommand \EOS [0]{\spacefactor3000\relax}%
\providecommand \BibitemShut  [1]{\csname bibitem#1\endcsname}%
\let\auto@bib@innerbib\@empty
\bibitem [{\citenamefont {La-O-Vorakiat}\ \emph {et~al.}(2009)\citenamefont {La-O-Vorakiat}, \citenamefont {Mathias}, \citenamefont {Grychtol}, \citenamefont {Adam}, \citenamefont {Siemens}, \citenamefont {Shaw}, \citenamefont {Nembach}, \citenamefont {Schneider}, \citenamefont {Aeschlimann}, \citenamefont {Silva} \emph {et~al.}}]{la2009ultrafast}%
  \BibitemOpen
  \bibfield  {author} {\bibinfo {author} {\bibfnamefont {C.}~\bibnamefont {La-O-Vorakiat}}, \bibinfo {author} {\bibfnamefont {S.}~\bibnamefont {Mathias}}, \bibinfo {author} {\bibfnamefont {P.}~\bibnamefont {Grychtol}}, \bibinfo {author} {\bibfnamefont {R.}~\bibnamefont {Adam}}, \bibinfo {author} {\bibfnamefont {M.}~\bibnamefont {Siemens}}, \bibinfo {author} {\bibfnamefont {J.}~\bibnamefont {Shaw}}, \bibinfo {author} {\bibfnamefont {H.}~\bibnamefont {Nembach}}, \bibinfo {author} {\bibfnamefont {C.}~\bibnamefont {Schneider}}, \bibinfo {author} {\bibfnamefont {M.}~\bibnamefont {Aeschlimann}}, \bibinfo {author} {\bibfnamefont {T.}~\bibnamefont {Silva}}, \emph {et~al.},\ }\bibfield  {title} {\bibinfo {title} {Ultrafast magnetization dynamics probed at elemental {M}-edges of {Ni} and {Fe} using tabletop high-order harmonic {EUV} light},\ }\href {https://journals.aps.org/prx/abstract/10.1103/PhysRevX.2.011005} {\bibfield  {journal} {\bibinfo  {journal} {Bulletin of the American Physical Society}\ }\textbf {\bibinfo
  {volume} {54}} (\bibinfo {year} {2009})}\BibitemShut {NoStop}%
\bibitem [{\citenamefont {La-O-Vorakiat}\ \emph {et~al.}(2012)\citenamefont {La-O-Vorakiat}, \citenamefont {Turgut}, \citenamefont {Teale}, \citenamefont {Kapteyn}, \citenamefont {Murnane}, \citenamefont {Mathias}, \citenamefont {Aeschlimann}, \citenamefont {Schneider}, \citenamefont {Shaw}, \citenamefont {Nembach},\ and\ \citenamefont {Silva}}]{la-o-vorakiat_ultrafast_2012}%
  \BibitemOpen
  \bibfield  {author} {\bibinfo {author} {\bibfnamefont {C.}~\bibnamefont {La-O-Vorakiat}}, \bibinfo {author} {\bibfnamefont {E.}~\bibnamefont {Turgut}}, \bibinfo {author} {\bibfnamefont {C.~A.}\ \bibnamefont {Teale}}, \bibinfo {author} {\bibfnamefont {H.~C.}\ \bibnamefont {Kapteyn}}, \bibinfo {author} {\bibfnamefont {M.~M.}\ \bibnamefont {Murnane}}, \bibinfo {author} {\bibfnamefont {S.}~\bibnamefont {Mathias}}, \bibinfo {author} {\bibfnamefont {M.}~\bibnamefont {Aeschlimann}}, \bibinfo {author} {\bibfnamefont {C.~M.}\ \bibnamefont {Schneider}}, \bibinfo {author} {\bibfnamefont {J.~M.}\ \bibnamefont {Shaw}}, \bibinfo {author} {\bibfnamefont {H.~T.}\ \bibnamefont {Nembach}},\ and\ \bibinfo {author} {\bibfnamefont {T.~J.}\ \bibnamefont {Silva}},\ }\bibfield  {title} {\bibinfo {title} {Ultrafast {Demagnetization} {Measurements} {Using} {Extreme} {Ultraviolet} {Light}: {Comparison} of {Electronic} and {Magnetic} {Contributions}},\ }\href {https://doi.org/10.1103/PhysRevX.2.011005} {\bibfield  {journal} {\bibinfo
  {journal} {Physical Review X}\ }\textbf {\bibinfo {volume} {2}},\ \bibinfo {pages} {011005} (\bibinfo {year} {2012})}\BibitemShut {NoStop}%
\bibitem [{\citenamefont {Mathias}\ \emph {et~al.}(2012)\citenamefont {Mathias}, \citenamefont {La-O-Vorakiat}, \citenamefont {Grychtol}, \citenamefont {Granitzka}, \citenamefont {Turgut}, \citenamefont {Shaw}, \citenamefont {Adam}, \citenamefont {Nembach}, \citenamefont {Siemens}, \citenamefont {Eich}, \citenamefont {Schneider}, \citenamefont {Silva}, \citenamefont {Aeschlimann}, \citenamefont {Murnane},\ and\ \citenamefont {Kapteyn}}]{mathias_probing_2012}%
  \BibitemOpen
  \bibfield  {author} {\bibinfo {author} {\bibfnamefont {S.}~\bibnamefont {Mathias}}, \bibinfo {author} {\bibfnamefont {C.}~\bibnamefont {La-O-Vorakiat}}, \bibinfo {author} {\bibfnamefont {P.}~\bibnamefont {Grychtol}}, \bibinfo {author} {\bibfnamefont {P.}~\bibnamefont {Granitzka}}, \bibinfo {author} {\bibfnamefont {E.}~\bibnamefont {Turgut}}, \bibinfo {author} {\bibfnamefont {J.~M.}\ \bibnamefont {Shaw}}, \bibinfo {author} {\bibfnamefont {R.}~\bibnamefont {Adam}}, \bibinfo {author} {\bibfnamefont {H.~T.}\ \bibnamefont {Nembach}}, \bibinfo {author} {\bibfnamefont {M.~E.}\ \bibnamefont {Siemens}}, \bibinfo {author} {\bibfnamefont {S.}~\bibnamefont {Eich}}, \bibinfo {author} {\bibfnamefont {C.~M.}\ \bibnamefont {Schneider}}, \bibinfo {author} {\bibfnamefont {T.~J.}\ \bibnamefont {Silva}}, \bibinfo {author} {\bibfnamefont {M.}~\bibnamefont {Aeschlimann}}, \bibinfo {author} {\bibfnamefont {M.~M.}\ \bibnamefont {Murnane}},\ and\ \bibinfo {author} {\bibfnamefont {H.~C.}\ \bibnamefont {Kapteyn}},\ }\bibfield
  {title} {\bibinfo {title} {Probing the timescale of the exchange interaction in a ferromagnetic alloy},\ }\href {https://doi.org/10.1073/pnas.1201371109} {\bibfield  {journal} {\bibinfo  {journal} {Proceedings of the National Academy of Sciences}\ }\textbf {\bibinfo {volume} {109}},\ \bibinfo {pages} {4792} (\bibinfo {year} {2012})}\BibitemShut {NoStop}%
\bibitem [{\citenamefont {Rudolf}\ \emph {et~al.}(2012)\citenamefont {Rudolf}, \citenamefont {La-O-Vorakiat}, \citenamefont {Battiato}, \citenamefont {Adam}, \citenamefont {Shaw}, \citenamefont {Turgut}, \citenamefont {Maldonado}, \citenamefont {Mathias}, \citenamefont {Grychtol}, \citenamefont {Nembach}, \citenamefont {Silva}, \citenamefont {Aeschlimann}, \citenamefont {Kapteyn}, \citenamefont {Murnane}, \citenamefont {Schneider},\ and\ \citenamefont {Oppeneer}}]{rudolf_ultrafast_2012}%
  \BibitemOpen
  \bibfield  {author} {\bibinfo {author} {\bibfnamefont {D.}~\bibnamefont {Rudolf}}, \bibinfo {author} {\bibfnamefont {C.}~\bibnamefont {La-O-Vorakiat}}, \bibinfo {author} {\bibfnamefont {M.}~\bibnamefont {Battiato}}, \bibinfo {author} {\bibfnamefont {R.}~\bibnamefont {Adam}}, \bibinfo {author} {\bibfnamefont {J.~M.}\ \bibnamefont {Shaw}}, \bibinfo {author} {\bibfnamefont {E.}~\bibnamefont {Turgut}}, \bibinfo {author} {\bibfnamefont {P.}~\bibnamefont {Maldonado}}, \bibinfo {author} {\bibfnamefont {S.}~\bibnamefont {Mathias}}, \bibinfo {author} {\bibfnamefont {P.}~\bibnamefont {Grychtol}}, \bibinfo {author} {\bibfnamefont {H.~T.}\ \bibnamefont {Nembach}}, \bibinfo {author} {\bibfnamefont {T.~J.}\ \bibnamefont {Silva}}, \bibinfo {author} {\bibfnamefont {M.}~\bibnamefont {Aeschlimann}}, \bibinfo {author} {\bibfnamefont {H.~C.}\ \bibnamefont {Kapteyn}}, \bibinfo {author} {\bibfnamefont {M.~M.}\ \bibnamefont {Murnane}}, \bibinfo {author} {\bibfnamefont {C.~M.}\ \bibnamefont {Schneider}},\ and\ \bibinfo {author}
  {\bibfnamefont {P.~M.}\ \bibnamefont {Oppeneer}},\ }\bibfield  {title} {\bibinfo {title} {Ultrafast magnetization enhancement in metallic multilayers driven by superdiffusive spin current},\ }\href {https://doi.org/10.1038/ncomms2029} {\bibfield  {journal} {\bibinfo  {journal} {Nature Communications}\ }\textbf {\bibinfo {volume} {3}},\ \bibinfo {pages} {1037} (\bibinfo {year} {2012})}\BibitemShut {NoStop}%
\bibitem [{\citenamefont {Turgut}\ \emph {et~al.}(2013)\citenamefont {Turgut}, \citenamefont {La-o vorakiat}, \citenamefont {Shaw}, \citenamefont {Grychtol}, \citenamefont {Nembach}, \citenamefont {Rudolf}, \citenamefont {Adam}, \citenamefont {Aeschlimann}, \citenamefont {Schneider}, \citenamefont {Silva}, \citenamefont {Murnane}, \citenamefont {Kapteyn},\ and\ \citenamefont {Mathias}}]{turgut_controlling_2013}%
  \BibitemOpen
  \bibfield  {author} {\bibinfo {author} {\bibfnamefont {E.}~\bibnamefont {Turgut}}, \bibinfo {author} {\bibfnamefont {C.}~\bibnamefont {La-o vorakiat}}, \bibinfo {author} {\bibfnamefont {J.~M.}\ \bibnamefont {Shaw}}, \bibinfo {author} {\bibfnamefont {P.}~\bibnamefont {Grychtol}}, \bibinfo {author} {\bibfnamefont {H.~T.}\ \bibnamefont {Nembach}}, \bibinfo {author} {\bibfnamefont {D.}~\bibnamefont {Rudolf}}, \bibinfo {author} {\bibfnamefont {R.}~\bibnamefont {Adam}}, \bibinfo {author} {\bibfnamefont {M.}~\bibnamefont {Aeschlimann}}, \bibinfo {author} {\bibfnamefont {C.~M.}\ \bibnamefont {Schneider}}, \bibinfo {author} {\bibfnamefont {T.~J.}\ \bibnamefont {Silva}}, \bibinfo {author} {\bibfnamefont {M.~M.}\ \bibnamefont {Murnane}}, \bibinfo {author} {\bibfnamefont {H.~C.}\ \bibnamefont {Kapteyn}},\ and\ \bibinfo {author} {\bibfnamefont {S.}~\bibnamefont {Mathias}},\ }\bibfield  {title} {\bibinfo {title} {Controlling the {Competition} between {Optically} {Induced} {Ultrafast} {Spin}-{Flip} {Scattering} and {Spin}
  {Transport} in {Magnetic} {Multilayers}},\ }\href {https://doi.org/10.1103/PhysRevLett.110.197201} {\bibfield  {journal} {\bibinfo  {journal} {Physical Review Letters}\ }\textbf {\bibinfo {volume} {110}},\ \bibinfo {pages} {197201} (\bibinfo {year} {2013})}\BibitemShut {NoStop}%
\bibitem [{\citenamefont {Tengdin}\ \emph {et~al.}(2020)\citenamefont {Tengdin}, \citenamefont {Gentry}, \citenamefont {Blonsky}, \citenamefont {Zusin}, \citenamefont {Gerrity}, \citenamefont {Hellbrück}, \citenamefont {Hofherr}, \citenamefont {Shaw}, \citenamefont {Kvashnin}, \citenamefont {Delczeg-Czirjak}, \citenamefont {Arora}, \citenamefont {Nembach}, \citenamefont {Silva}, \citenamefont {Mathias}, \citenamefont {Aeschlimann}, \citenamefont {Kapteyn}, \citenamefont {Thonig}, \citenamefont {Koumpouras}, \citenamefont {Eriksson},\ and\ \citenamefont {Murnane}}]{tengdin_direct_2020}%
  \BibitemOpen
  \bibfield  {author} {\bibinfo {author} {\bibfnamefont {P.}~\bibnamefont {Tengdin}}, \bibinfo {author} {\bibfnamefont {C.}~\bibnamefont {Gentry}}, \bibinfo {author} {\bibfnamefont {A.}~\bibnamefont {Blonsky}}, \bibinfo {author} {\bibfnamefont {D.}~\bibnamefont {Zusin}}, \bibinfo {author} {\bibfnamefont {M.}~\bibnamefont {Gerrity}}, \bibinfo {author} {\bibfnamefont {L.}~\bibnamefont {Hellbrück}}, \bibinfo {author} {\bibfnamefont {M.}~\bibnamefont {Hofherr}}, \bibinfo {author} {\bibfnamefont {J.}~\bibnamefont {Shaw}}, \bibinfo {author} {\bibfnamefont {Y.}~\bibnamefont {Kvashnin}}, \bibinfo {author} {\bibfnamefont {E.~K.}\ \bibnamefont {Delczeg-Czirjak}}, \bibinfo {author} {\bibfnamefont {M.}~\bibnamefont {Arora}}, \bibinfo {author} {\bibfnamefont {H.}~\bibnamefont {Nembach}}, \bibinfo {author} {\bibfnamefont {T.~J.}\ \bibnamefont {Silva}}, \bibinfo {author} {\bibfnamefont {S.}~\bibnamefont {Mathias}}, \bibinfo {author} {\bibfnamefont {M.}~\bibnamefont {Aeschlimann}}, \bibinfo {author} {\bibfnamefont {H.~C.}\
  \bibnamefont {Kapteyn}}, \bibinfo {author} {\bibfnamefont {D.}~\bibnamefont {Thonig}}, \bibinfo {author} {\bibfnamefont {K.}~\bibnamefont {Koumpouras}}, \bibinfo {author} {\bibfnamefont {O.}~\bibnamefont {Eriksson}},\ and\ \bibinfo {author} {\bibfnamefont {M.~M.}\ \bibnamefont {Murnane}},\ }\bibfield  {title} {\bibinfo {title} {Direct light–induced spin transfer between different elements in a spintronic {H}eusler material via femtosecond laser excitation},\ }\href {https://www.science.org/doi/full/10.1126/sciadv.aaz1100} {\bibfield  {journal} {\bibinfo  {journal} {Science Advances}\ }\textbf {\bibinfo {volume} {6}} (\bibinfo {year} {2020})}\BibitemShut {NoStop}%
\bibitem [{\citenamefont {von Korff~Schmising}\ \emph {et~al.}(2023)\citenamefont {von Korff~Schmising}, \citenamefont {Jana}, \citenamefont {Yao}, \citenamefont {Hennecke}, \citenamefont {Scheid}, \citenamefont {Sharma}, \citenamefont {Viret}, \citenamefont {Chauleau}, \citenamefont {Schick},\ and\ \citenamefont {Eisebitt}}]{KorffSchmising2023}%
  \BibitemOpen
  \bibfield  {author} {\bibinfo {author} {\bibfnamefont {C.}~\bibnamefont {von Korff~Schmising}}, \bibinfo {author} {\bibfnamefont {S.}~\bibnamefont {Jana}}, \bibinfo {author} {\bibfnamefont {K.}~\bibnamefont {Yao}}, \bibinfo {author} {\bibfnamefont {M.}~\bibnamefont {Hennecke}}, \bibinfo {author} {\bibfnamefont {P.}~\bibnamefont {Scheid}}, \bibinfo {author} {\bibfnamefont {S.}~\bibnamefont {Sharma}}, \bibinfo {author} {\bibfnamefont {M.}~\bibnamefont {Viret}}, \bibinfo {author} {\bibfnamefont {J.-Y.}\ \bibnamefont {Chauleau}}, \bibinfo {author} {\bibfnamefont {D.}~\bibnamefont {Schick}},\ and\ \bibinfo {author} {\bibfnamefont {S.}~\bibnamefont {Eisebitt}},\ }\bibfield  {title} {\bibinfo {title} {Ultrafast behavior of induced and intrinsic magnetic moments in {CoFeB}/{P}t bilayers probed by element-specific measurements in the extreme ultraviolet spectral range},\ }\href {https://doi.org/10.1103/physrevresearch.5.013147} {\bibfield  {journal} {\bibinfo  {journal} {Physical Review Research}\ }\textbf {\bibinfo
  {volume} {5}},\ \bibinfo {pages} {013147} (\bibinfo {year} {2023})}\BibitemShut {NoStop}%
\bibitem [{\citenamefont {Probst}\ \emph {et~al.}(2024)\citenamefont {Probst}, \citenamefont {Möller}, \citenamefont {Schumacher}, \citenamefont {Brede}, \citenamefont {Dewhurst}, \citenamefont {Reutzel}, \citenamefont {Steil}, \citenamefont {Sharma}, \citenamefont {Jansen},\ and\ \citenamefont {Mathias}}]{probst_unraveling_2024}%
  \BibitemOpen
  \bibfield  {author} {\bibinfo {author} {\bibfnamefont {H.}~\bibnamefont {Probst}}, \bibinfo {author} {\bibfnamefont {C.}~\bibnamefont {Möller}}, \bibinfo {author} {\bibfnamefont {M.}~\bibnamefont {Schumacher}}, \bibinfo {author} {\bibfnamefont {T.}~\bibnamefont {Brede}}, \bibinfo {author} {\bibfnamefont {J.~K.}\ \bibnamefont {Dewhurst}}, \bibinfo {author} {\bibfnamefont {M.}~\bibnamefont {Reutzel}}, \bibinfo {author} {\bibfnamefont {D.}~\bibnamefont {Steil}}, \bibinfo {author} {\bibfnamefont {S.}~\bibnamefont {Sharma}}, \bibinfo {author} {\bibfnamefont {G.~S.~M.}\ \bibnamefont {Jansen}},\ and\ \bibinfo {author} {\bibfnamefont {S.}~\bibnamefont {Mathias}},\ }\bibfield  {title} {\bibinfo {title} {Unraveling femtosecond spin and charge dynamics with extreme ultraviolet transverse {MOKE} spectroscopy},\ }\href {https://doi.org/10.1103/PhysRevResearch.6.013107} {\bibfield  {journal} {\bibinfo  {journal} {Physical Review Research}\ }\textbf {\bibinfo {volume} {6}},\ \bibinfo {pages} {013107} (\bibinfo {year}
  {2024})}\BibitemShut {NoStop}%
\bibitem [{\citenamefont {Gupta}\ \emph {et~al.}(2023)\citenamefont {Gupta}, \citenamefont {Cosco}, \citenamefont {Malik}, \citenamefont {Chen}, \citenamefont {Saha}, \citenamefont {Ghosh}, \citenamefont {Pohlmann}, \citenamefont {Mardegan}, \citenamefont {Francoual}, \citenamefont {Stefanuik}, \citenamefont {Söderström}, \citenamefont {Sanyal}, \citenamefont {Karis}, \citenamefont {Svedlindh}, \citenamefont {Oppeneer},\ and\ \citenamefont {Knut}}]{Gupta2023}%
  \BibitemOpen
  \bibfield  {author} {\bibinfo {author} {\bibfnamefont {R.}~\bibnamefont {Gupta}}, \bibinfo {author} {\bibfnamefont {F.}~\bibnamefont {Cosco}}, \bibinfo {author} {\bibfnamefont {R.~S.}\ \bibnamefont {Malik}}, \bibinfo {author} {\bibfnamefont {X.}~\bibnamefont {Chen}}, \bibinfo {author} {\bibfnamefont {S.}~\bibnamefont {Saha}}, \bibinfo {author} {\bibfnamefont {A.}~\bibnamefont {Ghosh}}, \bibinfo {author} {\bibfnamefont {T.}~\bibnamefont {Pohlmann}}, \bibinfo {author} {\bibfnamefont {J.~R.~L.}\ \bibnamefont {Mardegan}}, \bibinfo {author} {\bibfnamefont {S.}~\bibnamefont {Francoual}}, \bibinfo {author} {\bibfnamefont {R.}~\bibnamefont {Stefanuik}}, \bibinfo {author} {\bibfnamefont {J.}~\bibnamefont {Söderström}}, \bibinfo {author} {\bibfnamefont {B.}~\bibnamefont {Sanyal}}, \bibinfo {author} {\bibfnamefont {O.}~\bibnamefont {Karis}}, \bibinfo {author} {\bibfnamefont {P.}~\bibnamefont {Svedlindh}}, \bibinfo {author} {\bibfnamefont {P.~M.}\ \bibnamefont {Oppeneer}},\ and\ \bibinfo {author} {\bibfnamefont
  {R.}~\bibnamefont {Knut}},\ }\bibfield  {title} {\bibinfo {title} {Element-resolved evidence of superdiffusive spin current arising from ultrafast demagnetization process},\ }\href {https://doi.org/10.1103/physrevb.108.064427} {\bibfield  {journal} {\bibinfo  {journal} {Physical Review B}\ }\textbf {\bibinfo {volume} {108}},\ \bibinfo {pages} {064427} (\bibinfo {year} {2023})}\BibitemShut {NoStop}%
\bibitem [{\citenamefont {Turgut}\ \emph {et~al.}(2016)\citenamefont {Turgut}, \citenamefont {Zusin}, \citenamefont {Legut}, \citenamefont {Carva}, \citenamefont {Knut}, \citenamefont {Shaw}, \citenamefont {Chen}, \citenamefont {Tao}, \citenamefont {Nembach}, \citenamefont {Silva}, \citenamefont {Mathias}, \citenamefont {Aeschlimann}, \citenamefont {Oppeneer}, \citenamefont {Kapteyn}, \citenamefont {Murnane},\ and\ \citenamefont {Grychtol}}]{turgut_stoner_2016}%
  \BibitemOpen
  \bibfield  {author} {\bibinfo {author} {\bibfnamefont {E.}~\bibnamefont {Turgut}}, \bibinfo {author} {\bibfnamefont {D.}~\bibnamefont {Zusin}}, \bibinfo {author} {\bibfnamefont {D.}~\bibnamefont {Legut}}, \bibinfo {author} {\bibfnamefont {K.}~\bibnamefont {Carva}}, \bibinfo {author} {\bibfnamefont {R.}~\bibnamefont {Knut}}, \bibinfo {author} {\bibfnamefont {J.~M.}\ \bibnamefont {Shaw}}, \bibinfo {author} {\bibfnamefont {C.}~\bibnamefont {Chen}}, \bibinfo {author} {\bibfnamefont {Z.}~\bibnamefont {Tao}}, \bibinfo {author} {\bibfnamefont {H.~T.}\ \bibnamefont {Nembach}}, \bibinfo {author} {\bibfnamefont {T.~J.}\ \bibnamefont {Silva}}, \bibinfo {author} {\bibfnamefont {S.}~\bibnamefont {Mathias}}, \bibinfo {author} {\bibfnamefont {M.}~\bibnamefont {Aeschlimann}}, \bibinfo {author} {\bibfnamefont {P.~M.}\ \bibnamefont {Oppeneer}}, \bibinfo {author} {\bibfnamefont {H.~C.}\ \bibnamefont {Kapteyn}}, \bibinfo {author} {\bibfnamefont {M.~M.}\ \bibnamefont {Murnane}},\ and\ \bibinfo {author} {\bibfnamefont
  {P.}~\bibnamefont {Grychtol}},\ }\bibfield  {title} {\bibinfo {title} {Stoner versus {Heisenberg}: {Ultrafast} exchange reduction and magnon generation during laser-induced demagnetization},\ }\href {https://doi.org/10.1103/PhysRevB.94.220408} {\bibfield  {journal} {\bibinfo  {journal} {Physical Review B}\ }\textbf {\bibinfo {volume} {94}},\ \bibinfo {pages} {220408} (\bibinfo {year} {2016})}\BibitemShut {NoStop}%
\bibitem [{\citenamefont {Zusin}\ \emph {et~al.}(2018)\citenamefont {Zusin}, \citenamefont {Tengdin}, \citenamefont {Gopalakrishnan}, \citenamefont {Gentry}, \citenamefont {Blonsky}, \citenamefont {Gerrity}, \citenamefont {Legut}, \citenamefont {Shaw}, \citenamefont {Nembach}, \citenamefont {Silva}, \citenamefont {Oppeneer}, \citenamefont {Kapteyn},\ and\ \citenamefont {Murnane}}]{zusin_direct_2018}%
  \BibitemOpen
  \bibfield  {author} {\bibinfo {author} {\bibfnamefont {D.}~\bibnamefont {Zusin}}, \bibinfo {author} {\bibfnamefont {P.~M.}\ \bibnamefont {Tengdin}}, \bibinfo {author} {\bibfnamefont {M.}~\bibnamefont {Gopalakrishnan}}, \bibinfo {author} {\bibfnamefont {C.}~\bibnamefont {Gentry}}, \bibinfo {author} {\bibfnamefont {A.}~\bibnamefont {Blonsky}}, \bibinfo {author} {\bibfnamefont {M.}~\bibnamefont {Gerrity}}, \bibinfo {author} {\bibfnamefont {D.}~\bibnamefont {Legut}}, \bibinfo {author} {\bibfnamefont {J.~M.}\ \bibnamefont {Shaw}}, \bibinfo {author} {\bibfnamefont {H.~T.}\ \bibnamefont {Nembach}}, \bibinfo {author} {\bibfnamefont {T.~J.}\ \bibnamefont {Silva}}, \bibinfo {author} {\bibfnamefont {P.~M.}\ \bibnamefont {Oppeneer}}, \bibinfo {author} {\bibfnamefont {H.~C.}\ \bibnamefont {Kapteyn}},\ and\ \bibinfo {author} {\bibfnamefont {M.~M.}\ \bibnamefont {Murnane}},\ }\bibfield  {title} {\bibinfo {title} {Direct measurement of the static and transient magneto-optical permittivity of cobalt across the entire
  {M}-edge in reflection geometry by use of polarization scanning},\ }\href {https://doi.org/10.1103/PhysRevB.97.024433} {\bibfield  {journal} {\bibinfo  {journal} {Physical Review B}\ }\textbf {\bibinfo {volume} {97}},\ \bibinfo {pages} {024433} (\bibinfo {year} {2018})}\BibitemShut {NoStop}%
\bibitem [{\citenamefont {Hennecke}\ \emph {et~al.}(2022)\citenamefont {Hennecke}, \citenamefont {Schick}, \citenamefont {Sidiropoulos}, \citenamefont {Willems}, \citenamefont {Heilmann}, \citenamefont {Bock}, \citenamefont {Ehrentraut}, \citenamefont {Engel}, \citenamefont {Hessing}, \citenamefont {Pfau}, \citenamefont {Schmidbauer}, \citenamefont {Furchner}, \citenamefont {Schnuerer}, \citenamefont {von Korff~Schmising},\ and\ \citenamefont {Eisebitt}}]{Hennecke2022}%
  \BibitemOpen
  \bibfield  {author} {\bibinfo {author} {\bibfnamefont {M.}~\bibnamefont {Hennecke}}, \bibinfo {author} {\bibfnamefont {D.}~\bibnamefont {Schick}}, \bibinfo {author} {\bibfnamefont {T.}~\bibnamefont {Sidiropoulos}}, \bibinfo {author} {\bibfnamefont {F.}~\bibnamefont {Willems}}, \bibinfo {author} {\bibfnamefont {A.}~\bibnamefont {Heilmann}}, \bibinfo {author} {\bibfnamefont {M.}~\bibnamefont {Bock}}, \bibinfo {author} {\bibfnamefont {L.}~\bibnamefont {Ehrentraut}}, \bibinfo {author} {\bibfnamefont {D.}~\bibnamefont {Engel}}, \bibinfo {author} {\bibfnamefont {P.}~\bibnamefont {Hessing}}, \bibinfo {author} {\bibfnamefont {B.}~\bibnamefont {Pfau}}, \bibinfo {author} {\bibfnamefont {M.}~\bibnamefont {Schmidbauer}}, \bibinfo {author} {\bibfnamefont {A.}~\bibnamefont {Furchner}}, \bibinfo {author} {\bibfnamefont {M.}~\bibnamefont {Schnuerer}}, \bibinfo {author} {\bibfnamefont {C.}~\bibnamefont {von Korff~Schmising}},\ and\ \bibinfo {author} {\bibfnamefont {S.}~\bibnamefont {Eisebitt}},\ }\bibfield  {title}
  {\bibinfo {title} {Ultrafast element- and depth-resolved magnetization dynamics probed by transverse magneto-optical {K}err effect spectroscopy in the soft x-ray range},\ }\href {https://doi.org/10.1103/physrevresearch.4.l022062} {\bibfield  {journal} {\bibinfo  {journal} {Physical Review Research}\ }\textbf {\bibinfo {volume} {4}},\ \bibinfo {pages} {l022062} (\bibinfo {year} {2022})}\BibitemShut {NoStop}%
\bibitem [{\citenamefont {Chardonnet}\ \emph {et~al.}(2021)\citenamefont {Chardonnet}, \citenamefont {Hennes}, \citenamefont {Jarrier}, \citenamefont {Delaunay}, \citenamefont {Jaouen}, \citenamefont {Kuhlmann}, \citenamefont {Ekanayake}, \citenamefont {Léveillé}, \citenamefont {von Korff~Schmising}, \citenamefont {Schick}, \citenamefont {Yao}, \citenamefont {Liu}, \citenamefont {Chiuzbăian}, \citenamefont {Lüning}, \citenamefont {Vodungbo},\ and\ \citenamefont {Jal}}]{Chardonnet2021}%
  \BibitemOpen
  \bibfield  {author} {\bibinfo {author} {\bibfnamefont {V.}~\bibnamefont {Chardonnet}}, \bibinfo {author} {\bibfnamefont {M.}~\bibnamefont {Hennes}}, \bibinfo {author} {\bibfnamefont {R.}~\bibnamefont {Jarrier}}, \bibinfo {author} {\bibfnamefont {R.}~\bibnamefont {Delaunay}}, \bibinfo {author} {\bibfnamefont {N.}~\bibnamefont {Jaouen}}, \bibinfo {author} {\bibfnamefont {M.}~\bibnamefont {Kuhlmann}}, \bibinfo {author} {\bibfnamefont {N.}~\bibnamefont {Ekanayake}}, \bibinfo {author} {\bibfnamefont {C.}~\bibnamefont {Léveillé}}, \bibinfo {author} {\bibfnamefont {C.}~\bibnamefont {von Korff~Schmising}}, \bibinfo {author} {\bibfnamefont {D.}~\bibnamefont {Schick}}, \bibinfo {author} {\bibfnamefont {K.}~\bibnamefont {Yao}}, \bibinfo {author} {\bibfnamefont {X.}~\bibnamefont {Liu}}, \bibinfo {author} {\bibfnamefont {G.~S.}\ \bibnamefont {Chiuzbăian}}, \bibinfo {author} {\bibfnamefont {J.}~\bibnamefont {Lüning}}, \bibinfo {author} {\bibfnamefont {B.}~\bibnamefont {Vodungbo}},\ and\ \bibinfo {author}
  {\bibfnamefont {E.}~\bibnamefont {Jal}},\ }\bibfield  {title} {\bibinfo {title} {Toward ultrafast magnetic depth profiling using time-resolved x-ray resonant magnetic reflectivity},\ }\href {https://pmc.ncbi.nlm.nih.gov/articles/PMC8225393/} {\bibfield  {journal} {\bibinfo  {journal} {Structural Dynamics}\ }\textbf {\bibinfo {volume} {8}} (\bibinfo {year} {2021})}\BibitemShut {NoStop}%
\bibitem [{\citenamefont {Richter}\ \emph {et~al.}(2024)\citenamefont {Richter}, \citenamefont {Jana}, \citenamefont {Hennecke}, \citenamefont {Schick}, \citenamefont {von Korff~Schmising},\ and\ \citenamefont {Eisebitt}}]{Richter2024}%
  \BibitemOpen
  \bibfield  {author} {\bibinfo {author} {\bibfnamefont {J.}~\bibnamefont {Richter}}, \bibinfo {author} {\bibfnamefont {S.}~\bibnamefont {Jana}}, \bibinfo {author} {\bibfnamefont {M.}~\bibnamefont {Hennecke}}, \bibinfo {author} {\bibfnamefont {D.}~\bibnamefont {Schick}}, \bibinfo {author} {\bibfnamefont {C.}~\bibnamefont {von Korff~Schmising}},\ and\ \bibinfo {author} {\bibfnamefont {S.}~\bibnamefont {Eisebitt}},\ }\bibfield  {title} {\bibinfo {title} {Relationship between magnetic asymmetry and magnetization in ultrafast transverse magneto-optical {Kerr} effect spectroscopy in the extreme ultraviolet spectral range},\ }\href {https://doi.org/10.1103/PhysRevB.109.184440} {\bibfield  {journal} {\bibinfo  {journal} {Phys. Rev. B}\ }\textbf {\bibinfo {volume} {109}},\ \bibinfo {pages} {184440} (\bibinfo {year} {2024})}\BibitemShut {NoStop}%
\bibitem [{\citenamefont {Jana}\ \emph {et~al.}(2020)\citenamefont {Jana}, \citenamefont {Malik}, \citenamefont {Kvashnin}, \citenamefont {Locht}, \citenamefont {Knut}, \citenamefont {Stefanuik}, \citenamefont {Di~Marco}, \citenamefont {Yaresko}, \citenamefont {Ahlberg}, \citenamefont {Åkerman}, \citenamefont {Chimata}, \citenamefont {Battiato}, \citenamefont {Söderström}, \citenamefont {Eriksson},\ and\ \citenamefont {Karis}}]{jana_analysis_2020}%
  \BibitemOpen
  \bibfield  {author} {\bibinfo {author} {\bibfnamefont {S.}~\bibnamefont {Jana}}, \bibinfo {author} {\bibfnamefont {R.~S.}\ \bibnamefont {Malik}}, \bibinfo {author} {\bibfnamefont {Y.~O.}\ \bibnamefont {Kvashnin}}, \bibinfo {author} {\bibfnamefont {I.~L.~M.}\ \bibnamefont {Locht}}, \bibinfo {author} {\bibfnamefont {R.}~\bibnamefont {Knut}}, \bibinfo {author} {\bibfnamefont {R.}~\bibnamefont {Stefanuik}}, \bibinfo {author} {\bibfnamefont {I.}~\bibnamefont {Di~Marco}}, \bibinfo {author} {\bibfnamefont {A.~N.}\ \bibnamefont {Yaresko}}, \bibinfo {author} {\bibfnamefont {M.}~\bibnamefont {Ahlberg}}, \bibinfo {author} {\bibfnamefont {J.}~\bibnamefont {Åkerman}}, \bibinfo {author} {\bibfnamefont {R.}~\bibnamefont {Chimata}}, \bibinfo {author} {\bibfnamefont {M.}~\bibnamefont {Battiato}}, \bibinfo {author} {\bibfnamefont {J.}~\bibnamefont {Söderström}}, \bibinfo {author} {\bibfnamefont {O.}~\bibnamefont {Eriksson}},\ and\ \bibinfo {author} {\bibfnamefont {O.}~\bibnamefont {Karis}},\ }\bibfield  {title} {\bibinfo
  {title} {Analysis of the linear relationship between asymmetry and magnetic moment at the {M} edge of 3d transition metals},\ }\href {https://doi.org/10.1103/PhysRevResearch.2.013180} {\bibfield  {journal} {\bibinfo  {journal} {Physical Review Research}\ }\textbf {\bibinfo {volume} {2}},\ \bibinfo {pages} {013180} (\bibinfo {year} {2020})}\BibitemShut {NoStop}%
\bibitem [{\citenamefont {Alves}\ \emph {et~al.}(2019)\citenamefont {Alves}, \citenamefont {Lambert}, \citenamefont {Malka}, \citenamefont {Hehn}, \citenamefont {Malinowski}, \citenamefont {Hennes}, \citenamefont {Chardonnet}, \citenamefont {Jal}, \citenamefont {Lüning},\ and\ \citenamefont {Vodungbo}}]{Alves2019}%
  \BibitemOpen
  \bibfield  {author} {\bibinfo {author} {\bibfnamefont {C.}~\bibnamefont {Alves}}, \bibinfo {author} {\bibfnamefont {G.}~\bibnamefont {Lambert}}, \bibinfo {author} {\bibfnamefont {V.}~\bibnamefont {Malka}}, \bibinfo {author} {\bibfnamefont {M.}~\bibnamefont {Hehn}}, \bibinfo {author} {\bibfnamefont {G.}~\bibnamefont {Malinowski}}, \bibinfo {author} {\bibfnamefont {M.}~\bibnamefont {Hennes}}, \bibinfo {author} {\bibfnamefont {V.}~\bibnamefont {Chardonnet}}, \bibinfo {author} {\bibfnamefont {E.}~\bibnamefont {Jal}}, \bibinfo {author} {\bibfnamefont {J.}~\bibnamefont {Lüning}},\ and\ \bibinfo {author} {\bibfnamefont {B.}~\bibnamefont {Vodungbo}},\ }\bibfield  {title} {\bibinfo {title} {Resonant {Faraday} effect using high-order harmonics for the investigation of ultrafast demagnetization},\ }\href {https://doi.org/10.1103/physrevb.100.144421} {\bibfield  {journal} {\bibinfo  {journal} {Physical Review B}\ }\textbf {\bibinfo {volume} {100}},\ \bibinfo {pages} {144421} (\bibinfo {year} {2019})}\BibitemShut
  {NoStop}%
\bibitem [{\citenamefont {Yamamoto}\ \emph {et~al.}(2015)\citenamefont {Yamamoto}, \citenamefont {Taguchi}, \citenamefont {Someya}, \citenamefont {Kubota}, \citenamefont {Ito}, \citenamefont {Wadati}, \citenamefont {Fujisawa}, \citenamefont {Capotondi}, \citenamefont {Pedersoli}, \citenamefont {Manfredda}, \citenamefont {Raimondi}, \citenamefont {Kiskinova}, \citenamefont {Fujii}, \citenamefont {Moras}, \citenamefont {Tsuyama}, \citenamefont {Nakamura}, \citenamefont {Kato}, \citenamefont {Higashide}, \citenamefont {Iwata}, \citenamefont {Yamamoto}, \citenamefont {Shin},\ and\ \citenamefont {Matsuda}}]{yamamoto_ultrafast_2015}%
  \BibitemOpen
  \bibfield  {author} {\bibinfo {author} {\bibfnamefont {S.}~\bibnamefont {Yamamoto}}, \bibinfo {author} {\bibfnamefont {M.}~\bibnamefont {Taguchi}}, \bibinfo {author} {\bibfnamefont {T.}~\bibnamefont {Someya}}, \bibinfo {author} {\bibfnamefont {Y.}~\bibnamefont {Kubota}}, \bibinfo {author} {\bibfnamefont {S.}~\bibnamefont {Ito}}, \bibinfo {author} {\bibfnamefont {H.}~\bibnamefont {Wadati}}, \bibinfo {author} {\bibfnamefont {M.}~\bibnamefont {Fujisawa}}, \bibinfo {author} {\bibfnamefont {F.}~\bibnamefont {Capotondi}}, \bibinfo {author} {\bibfnamefont {E.}~\bibnamefont {Pedersoli}}, \bibinfo {author} {\bibfnamefont {M.}~\bibnamefont {Manfredda}}, \bibinfo {author} {\bibfnamefont {L.}~\bibnamefont {Raimondi}}, \bibinfo {author} {\bibfnamefont {M.}~\bibnamefont {Kiskinova}}, \bibinfo {author} {\bibfnamefont {J.}~\bibnamefont {Fujii}}, \bibinfo {author} {\bibfnamefont {P.}~\bibnamefont {Moras}}, \bibinfo {author} {\bibfnamefont {T.}~\bibnamefont {Tsuyama}}, \bibinfo {author} {\bibfnamefont {T.}~\bibnamefont
  {Nakamura}}, \bibinfo {author} {\bibfnamefont {T.}~\bibnamefont {Kato}}, \bibinfo {author} {\bibfnamefont {T.}~\bibnamefont {Higashide}}, \bibinfo {author} {\bibfnamefont {S.}~\bibnamefont {Iwata}}, \bibinfo {author} {\bibfnamefont {S.}~\bibnamefont {Yamamoto}}, \bibinfo {author} {\bibfnamefont {S.}~\bibnamefont {Shin}},\ and\ \bibinfo {author} {\bibfnamefont {I.}~\bibnamefont {Matsuda}},\ }\bibfield  {title} {\bibinfo {title} {Ultrafast spin-switching of a ferrimagnetic alloy at room temperature traced by resonant magneto-optical {Kerr} effect using a seeded free electron laser},\ }\href {https://doi.org/10.1063/1.4927828} {\bibfield  {journal} {\bibinfo  {journal} {Review of Scientific Instruments}\ }\textbf {\bibinfo {volume} {86}},\ \bibinfo {pages} {083901} (\bibinfo {year} {2015})}\BibitemShut {NoStop}%
\bibitem [{\citenamefont {Yamamoto}\ \emph {et~al.}(2020)\citenamefont {Yamamoto}, \citenamefont {Moussaoui}, \citenamefont {Hirata}, \citenamefont {Yamamoto}, \citenamefont {Kubota}, \citenamefont {Owada}, \citenamefont {Yabashi}, \citenamefont {Seki}, \citenamefont {Takanashi}, \citenamefont {Matsuda},\ and\ \citenamefont {Wadati}}]{Yamamoto2020}%
  \BibitemOpen
  \bibfield  {author} {\bibinfo {author} {\bibfnamefont {K.}~\bibnamefont {Yamamoto}}, \bibinfo {author} {\bibfnamefont {S.~E.}\ \bibnamefont {Moussaoui}}, \bibinfo {author} {\bibfnamefont {Y.}~\bibnamefont {Hirata}}, \bibinfo {author} {\bibfnamefont {S.}~\bibnamefont {Yamamoto}}, \bibinfo {author} {\bibfnamefont {Y.}~\bibnamefont {Kubota}}, \bibinfo {author} {\bibfnamefont {S.}~\bibnamefont {Owada}}, \bibinfo {author} {\bibfnamefont {M.}~\bibnamefont {Yabashi}}, \bibinfo {author} {\bibfnamefont {T.}~\bibnamefont {Seki}}, \bibinfo {author} {\bibfnamefont {K.}~\bibnamefont {Takanashi}}, \bibinfo {author} {\bibfnamefont {I.}~\bibnamefont {Matsuda}},\ and\ \bibinfo {author} {\bibfnamefont {H.}~\bibnamefont {Wadati}},\ }\bibfield  {title} {\bibinfo {title} {Element-selectively tracking ultrafast demagnetization process in {Co}/{Pt} multilayer thin films by the resonant magneto-optical {Kerr} effect},\ }\href {https://pubs.aip.org/aip/apl/article/116/17/172406/38344/Element-selectively-tracking-ultrafast}
  {\bibfield  {journal} {\bibinfo  {journal} {Applied Physics Letters}\ }\textbf {\bibinfo {volume} {116}} (\bibinfo {year} {2020})}\BibitemShut {NoStop}%
\bibitem [{\citenamefont {von Korff~Schmising}\ \emph {et~al.}(2020)\citenamefont {von Korff~Schmising}, \citenamefont {Willems}, \citenamefont {Sharma}, \citenamefont {Yao}, \citenamefont {Borchert}, \citenamefont {Hennecke}, \citenamefont {Schick}, \citenamefont {Radu}, \citenamefont {Strüber}, \citenamefont {Engel}, \citenamefont {Shokeen}, \citenamefont {Buck}, \citenamefont {Bagschik}, \citenamefont {Viefhaus}, \citenamefont {Hartmann}, \citenamefont {Manschwetus}, \citenamefont {Grunewald}, \citenamefont {Düsterer}, \citenamefont {Jal}, \citenamefont {Vodungbo}, \citenamefont {Lüning},\ and\ \citenamefont {Eisebitt}}]{KorffSchmising2020}%
  \BibitemOpen
  \bibfield  {author} {\bibinfo {author} {\bibfnamefont {C.}~\bibnamefont {von Korff~Schmising}}, \bibinfo {author} {\bibfnamefont {F.}~\bibnamefont {Willems}}, \bibinfo {author} {\bibfnamefont {S.}~\bibnamefont {Sharma}}, \bibinfo {author} {\bibfnamefont {K.}~\bibnamefont {Yao}}, \bibinfo {author} {\bibfnamefont {M.}~\bibnamefont {Borchert}}, \bibinfo {author} {\bibfnamefont {M.}~\bibnamefont {Hennecke}}, \bibinfo {author} {\bibfnamefont {D.}~\bibnamefont {Schick}}, \bibinfo {author} {\bibfnamefont {I.}~\bibnamefont {Radu}}, \bibinfo {author} {\bibfnamefont {C.}~\bibnamefont {Strüber}}, \bibinfo {author} {\bibfnamefont {D.~W.}\ \bibnamefont {Engel}}, \bibinfo {author} {\bibfnamefont {V.}~\bibnamefont {Shokeen}}, \bibinfo {author} {\bibfnamefont {J.}~\bibnamefont {Buck}}, \bibinfo {author} {\bibfnamefont {K.}~\bibnamefont {Bagschik}}, \bibinfo {author} {\bibfnamefont {J.}~\bibnamefont {Viefhaus}}, \bibinfo {author} {\bibfnamefont {G.}~\bibnamefont {Hartmann}}, \bibinfo {author} {\bibfnamefont
  {B.}~\bibnamefont {Manschwetus}}, \bibinfo {author} {\bibfnamefont {S.}~\bibnamefont {Grunewald}}, \bibinfo {author} {\bibfnamefont {S.}~\bibnamefont {Düsterer}}, \bibinfo {author} {\bibfnamefont {E.}~\bibnamefont {Jal}}, \bibinfo {author} {\bibfnamefont {B.}~\bibnamefont {Vodungbo}}, \bibinfo {author} {\bibfnamefont {J.}~\bibnamefont {Lüning}},\ and\ \bibinfo {author} {\bibfnamefont {S.}~\bibnamefont {Eisebitt}},\ }\bibfield  {title} {\bibinfo {title} {Element-specific magnetization dynamics of complex magnetic systems probed by ultrafast magneto-optical spectroscopy},\ }\href {https://doi.org/10.3390/app10217580} {\bibfield  {journal} {\bibinfo  {journal} {Applied Sciences}\ }\textbf {\bibinfo {volume} {10}},\ \bibinfo {pages} {7580} (\bibinfo {year} {2020})}\BibitemShut {NoStop}%
\bibitem [{\citenamefont {Caretta}\ \emph {et~al.}(2021)\citenamefont {Caretta}, \citenamefont {Laterza}, \citenamefont {Bonanni}, \citenamefont {Sergo}, \citenamefont {Dri}, \citenamefont {Cautero}, \citenamefont {Galassi}, \citenamefont {Zamolo}, \citenamefont {Simoncig}, \citenamefont {Zangrando}, \citenamefont {Gessini}, \citenamefont {Zilio}, \citenamefont {Flammini}, \citenamefont {Moras}, \citenamefont {Demidovich}, \citenamefont {Danailov}, \citenamefont {Parmigiani},\ and\ \citenamefont {Malvestuto}}]{caretta_novel_2021}%
  \BibitemOpen
  \bibfield  {author} {\bibinfo {author} {\bibfnamefont {A.}~\bibnamefont {Caretta}}, \bibinfo {author} {\bibfnamefont {S.}~\bibnamefont {Laterza}}, \bibinfo {author} {\bibfnamefont {V.}~\bibnamefont {Bonanni}}, \bibinfo {author} {\bibfnamefont {R.}~\bibnamefont {Sergo}}, \bibinfo {author} {\bibfnamefont {C.}~\bibnamefont {Dri}}, \bibinfo {author} {\bibfnamefont {G.}~\bibnamefont {Cautero}}, \bibinfo {author} {\bibfnamefont {F.}~\bibnamefont {Galassi}}, \bibinfo {author} {\bibfnamefont {M.}~\bibnamefont {Zamolo}}, \bibinfo {author} {\bibfnamefont {A.}~\bibnamefont {Simoncig}}, \bibinfo {author} {\bibfnamefont {M.}~\bibnamefont {Zangrando}}, \bibinfo {author} {\bibfnamefont {A.}~\bibnamefont {Gessini}}, \bibinfo {author} {\bibfnamefont {S.~D.}\ \bibnamefont {Zilio}}, \bibinfo {author} {\bibfnamefont {R.}~\bibnamefont {Flammini}}, \bibinfo {author} {\bibfnamefont {P.}~\bibnamefont {Moras}}, \bibinfo {author} {\bibfnamefont {A.}~\bibnamefont {Demidovich}}, \bibinfo {author} {\bibfnamefont {M.}~\bibnamefont
  {Danailov}}, \bibinfo {author} {\bibfnamefont {F.}~\bibnamefont {Parmigiani}},\ and\ \bibinfo {author} {\bibfnamefont {M.}~\bibnamefont {Malvestuto}},\ }\bibfield  {title} {\bibinfo {title} {A novel free-electron laser single-pulse {Wollaston} polarimeter for magneto-dynamical studies},\ }\href {https://doi.org/10.1063/4.0000104} {\bibfield  {journal} {\bibinfo  {journal} {Structural Dynamics}\ }\textbf {\bibinfo {volume} {8}},\ \bibinfo {pages} {034304} (\bibinfo {year} {2021})}\BibitemShut {NoStop}%
\bibitem [{\citenamefont {Pancaldi}\ \emph {et~al.}(2022)\citenamefont {Pancaldi}, \citenamefont {Strüber}, \citenamefont {Friedrich}, \citenamefont {Pedersoli}, \citenamefont {De~Angelis}, \citenamefont {Nikolov}, \citenamefont {Manfredda}, \citenamefont {Foglia}, \citenamefont {Yulin}, \citenamefont {Spezzani}, \citenamefont {Sacchi}, \citenamefont {Eisebitt}, \citenamefont {Von Korff~Schmising},\ and\ \citenamefont {Capotondi}}]{pancaldi_comix_2022}%
  \BibitemOpen
  \bibfield  {author} {\bibinfo {author} {\bibfnamefont {M.}~\bibnamefont {Pancaldi}}, \bibinfo {author} {\bibfnamefont {C.}~\bibnamefont {Strüber}}, \bibinfo {author} {\bibfnamefont {B.}~\bibnamefont {Friedrich}}, \bibinfo {author} {\bibfnamefont {E.}~\bibnamefont {Pedersoli}}, \bibinfo {author} {\bibfnamefont {D.}~\bibnamefont {De~Angelis}}, \bibinfo {author} {\bibfnamefont {I.~P.}\ \bibnamefont {Nikolov}}, \bibinfo {author} {\bibfnamefont {M.}~\bibnamefont {Manfredda}}, \bibinfo {author} {\bibfnamefont {L.}~\bibnamefont {Foglia}}, \bibinfo {author} {\bibfnamefont {S.}~\bibnamefont {Yulin}}, \bibinfo {author} {\bibfnamefont {C.}~\bibnamefont {Spezzani}}, \bibinfo {author} {\bibfnamefont {M.}~\bibnamefont {Sacchi}}, \bibinfo {author} {\bibfnamefont {S.}~\bibnamefont {Eisebitt}}, \bibinfo {author} {\bibfnamefont {C.}~\bibnamefont {Von Korff~Schmising}},\ and\ \bibinfo {author} {\bibfnamefont {F.}~\bibnamefont {Capotondi}},\ }\bibfield  {title} {\bibinfo {title} {The {COMIX} polarimeter: a compact device for
  {XUV} polarization analysis},\ }\href {https://doi.org/10.1107/S1600577522004027} {\bibfield  {journal} {\bibinfo  {journal} {Journal of Synchrotron Radiation}\ }\textbf {\bibinfo {volume} {29}},\ \bibinfo {pages} {969} (\bibinfo {year} {2022})}\BibitemShut {NoStop}%
\bibitem [{\citenamefont {Willems}\ \emph {et~al.}(2015)\citenamefont {Willems}, \citenamefont {Smeenk}, \citenamefont {Zhavoronkov}, \citenamefont {Kornilov}, \citenamefont {Radu}, \citenamefont {Schmidbauer}, \citenamefont {Hanke}, \citenamefont {von Korff~Schmising}, \citenamefont {Vrakking},\ and\ \citenamefont {Eisebitt}}]{willems_probing_2015}%
  \BibitemOpen
  \bibfield  {author} {\bibinfo {author} {\bibfnamefont {F.}~\bibnamefont {Willems}}, \bibinfo {author} {\bibfnamefont {C.~T.~L.}\ \bibnamefont {Smeenk}}, \bibinfo {author} {\bibfnamefont {N.}~\bibnamefont {Zhavoronkov}}, \bibinfo {author} {\bibfnamefont {O.}~\bibnamefont {Kornilov}}, \bibinfo {author} {\bibfnamefont {I.}~\bibnamefont {Radu}}, \bibinfo {author} {\bibfnamefont {M.}~\bibnamefont {Schmidbauer}}, \bibinfo {author} {\bibfnamefont {M.}~\bibnamefont {Hanke}}, \bibinfo {author} {\bibfnamefont {C.}~\bibnamefont {von Korff~Schmising}}, \bibinfo {author} {\bibfnamefont {M.~J.~J.}\ \bibnamefont {Vrakking}},\ and\ \bibinfo {author} {\bibfnamefont {S.}~\bibnamefont {Eisebitt}},\ }\bibfield  {title} {\bibinfo {title} {Probing ultrafast spin dynamics with high-harmonic magnetic circular dichroism spectroscopy},\ }\href {https://doi.org/10.1103/PhysRevB.92.220405} {\bibfield  {journal} {\bibinfo  {journal} {Physical Review B}\ }\textbf {\bibinfo {volume} {92}},\ \bibinfo {pages} {220405} (\bibinfo {year}
  {2015})}\BibitemShut {NoStop}%
\bibitem [{\citenamefont {Siegrist}\ \emph {et~al.}(2019)\citenamefont {Siegrist}, \citenamefont {Gessner}, \citenamefont {Ossiander}, \citenamefont {Denker}, \citenamefont {Chang}, \citenamefont {Schröder}, \citenamefont {Guggenmos}, \citenamefont {Cui}, \citenamefont {Walowski}, \citenamefont {Martens}, \citenamefont {Dewhurst}, \citenamefont {Kleineberg}, \citenamefont {Münzenberg}, \citenamefont {Sharma},\ and\ \citenamefont {Schultze}}]{Siegrist2019}%
  \BibitemOpen
  \bibfield  {author} {\bibinfo {author} {\bibfnamefont {F.}~\bibnamefont {Siegrist}}, \bibinfo {author} {\bibfnamefont {J.~A.}\ \bibnamefont {Gessner}}, \bibinfo {author} {\bibfnamefont {M.}~\bibnamefont {Ossiander}}, \bibinfo {author} {\bibfnamefont {C.}~\bibnamefont {Denker}}, \bibinfo {author} {\bibfnamefont {Y.-P.}\ \bibnamefont {Chang}}, \bibinfo {author} {\bibfnamefont {M.~C.}\ \bibnamefont {Schröder}}, \bibinfo {author} {\bibfnamefont {A.}~\bibnamefont {Guggenmos}}, \bibinfo {author} {\bibfnamefont {Y.}~\bibnamefont {Cui}}, \bibinfo {author} {\bibfnamefont {J.}~\bibnamefont {Walowski}}, \bibinfo {author} {\bibfnamefont {U.}~\bibnamefont {Martens}}, \bibinfo {author} {\bibfnamefont {J.~K.}\ \bibnamefont {Dewhurst}}, \bibinfo {author} {\bibfnamefont {U.}~\bibnamefont {Kleineberg}}, \bibinfo {author} {\bibfnamefont {M.}~\bibnamefont {Münzenberg}}, \bibinfo {author} {\bibfnamefont {S.}~\bibnamefont {Sharma}},\ and\ \bibinfo {author} {\bibfnamefont {M.}~\bibnamefont {Schultze}},\ }\bibfield  {title}
  {\bibinfo {title} {Light-wave dynamic control of magnetism},\ }\href {https://doi.org/10.1038/s41586-019-1333-x} {\bibfield  {journal} {\bibinfo  {journal} {Nature}\ }\textbf {\bibinfo {volume} {571}},\ \bibinfo {pages} {240} (\bibinfo {year} {2019})}\BibitemShut {NoStop}%
\bibitem [{\citenamefont {Willems}\ \emph {et~al.}(2020)\citenamefont {Willems}, \citenamefont {von Korff~Schmising}, \citenamefont {Strüber}, \citenamefont {Schick}, \citenamefont {Engel}, \citenamefont {Dewhurst}, \citenamefont {Elliott}, \citenamefont {Sharma},\ and\ \citenamefont {Eisebitt}}]{willems_optical_2020}%
  \BibitemOpen
  \bibfield  {author} {\bibinfo {author} {\bibfnamefont {F.}~\bibnamefont {Willems}}, \bibinfo {author} {\bibfnamefont {C.}~\bibnamefont {von Korff~Schmising}}, \bibinfo {author} {\bibfnamefont {C.}~\bibnamefont {Strüber}}, \bibinfo {author} {\bibfnamefont {D.}~\bibnamefont {Schick}}, \bibinfo {author} {\bibfnamefont {D.~W.}\ \bibnamefont {Engel}}, \bibinfo {author} {\bibfnamefont {J.~K.}\ \bibnamefont {Dewhurst}}, \bibinfo {author} {\bibfnamefont {P.}~\bibnamefont {Elliott}}, \bibinfo {author} {\bibfnamefont {S.}~\bibnamefont {Sharma}},\ and\ \bibinfo {author} {\bibfnamefont {S.}~\bibnamefont {Eisebitt}},\ }\bibfield  {title} {\bibinfo {title} {Optical inter-site spin transfer probed by energy and spin-resolved transient absorption spectroscopy},\ }\href {https://doi.org/10.1038/s41467-020-14691-5} {\bibfield  {journal} {\bibinfo  {journal} {Nature Communications}\ }\textbf {\bibinfo {volume} {11}},\ \bibinfo {pages} {871} (\bibinfo {year} {2020})}\BibitemShut {NoStop}%
\bibitem [{\citenamefont {Yao}\ \emph {et~al.}(2020{\natexlab{a}})\citenamefont {Yao}, \citenamefont {Willems}, \citenamefont {von Korff~Schmising}, \citenamefont {Radu}, \citenamefont {Str\"uber}, \citenamefont {Schick}, \citenamefont {Engel}, \citenamefont {Tsukamoto}, \citenamefont {Dewhurst}, \citenamefont {Sharma},\ and\ \citenamefont {Eisebitt}}]{Yao2020_a}%
  \BibitemOpen
  \bibfield  {author} {\bibinfo {author} {\bibfnamefont {K.}~\bibnamefont {Yao}}, \bibinfo {author} {\bibfnamefont {F.}~\bibnamefont {Willems}}, \bibinfo {author} {\bibfnamefont {C.}~\bibnamefont {von Korff~Schmising}}, \bibinfo {author} {\bibfnamefont {I.}~\bibnamefont {Radu}}, \bibinfo {author} {\bibfnamefont {C.}~\bibnamefont {Str\"uber}}, \bibinfo {author} {\bibfnamefont {D.}~\bibnamefont {Schick}}, \bibinfo {author} {\bibfnamefont {D.}~\bibnamefont {Engel}}, \bibinfo {author} {\bibfnamefont {A.}~\bibnamefont {Tsukamoto}}, \bibinfo {author} {\bibfnamefont {J.~K.}\ \bibnamefont {Dewhurst}}, \bibinfo {author} {\bibfnamefont {S.}~\bibnamefont {Sharma}},\ and\ \bibinfo {author} {\bibfnamefont {S.}~\bibnamefont {Eisebitt}},\ }\bibfield  {title} {\bibinfo {title} {Distinct spectral response in {M}-edge magnetic circular dichroism},\ }\href {https://link.aps.org/doi/10.1103/PhysRevB.102.100405} {\bibfield  {journal} {\bibinfo  {journal} {Phys. Rev. B}\ }\textbf {\bibinfo {volume} {102}},\ \bibinfo {pages}
  {100405} (\bibinfo {year} {2020}{\natexlab{a}})}\BibitemShut {NoStop}%
\bibitem [{\citenamefont {G\'eneaux}\ \emph {et~al.}(2024)\citenamefont {G\'eneaux}, \citenamefont {Chang}, \citenamefont {Guggenmos}, \citenamefont {Delaunay}, \citenamefont {L\'egar\'e}, \citenamefont {L\'egar\'e}, \citenamefont {L\"uning}, \citenamefont {Parpiiev}, \citenamefont {Molesky}, \citenamefont {de~Roulet}, \citenamefont {Zuerch}, \citenamefont {Sharma}, \citenamefont {Schultze},\ and\ \citenamefont {Leone}}]{Geneaux2024}%
  \BibitemOpen
  \bibfield  {author} {\bibinfo {author} {\bibfnamefont {R.}~\bibnamefont {G\'eneaux}}, \bibinfo {author} {\bibfnamefont {H.-T.}\ \bibnamefont {Chang}}, \bibinfo {author} {\bibfnamefont {A.}~\bibnamefont {Guggenmos}}, \bibinfo {author} {\bibfnamefont {R.}~\bibnamefont {Delaunay}}, \bibinfo {author} {\bibfnamefont {F.~m.~c.}\ \bibnamefont {L\'egar\'e}}, \bibinfo {author} {\bibfnamefont {K.}~\bibnamefont {L\'egar\'e}}, \bibinfo {author} {\bibfnamefont {J.}~\bibnamefont {L\"uning}}, \bibinfo {author} {\bibfnamefont {T.}~\bibnamefont {Parpiiev}}, \bibinfo {author} {\bibfnamefont {I.~J.~P.}\ \bibnamefont {Molesky}}, \bibinfo {author} {\bibfnamefont {B.~R.}\ \bibnamefont {de~Roulet}}, \bibinfo {author} {\bibfnamefont {M.~W.}\ \bibnamefont {Zuerch}}, \bibinfo {author} {\bibfnamefont {S.}~\bibnamefont {Sharma}}, \bibinfo {author} {\bibfnamefont {M.}~\bibnamefont {Schultze}},\ and\ \bibinfo {author} {\bibfnamefont {S.~R.}\ \bibnamefont {Leone}},\ }\bibfield  {title} {\bibinfo {title} {Spin dynamics across metallic
  layers on the few-femtosecond timescale},\ }\href {https://doi.org/10.1103/PhysRevLett.133.106902} {\bibfield  {journal} {\bibinfo  {journal} {Phys. Rev. Lett.}\ }\textbf {\bibinfo {volume} {133}},\ \bibinfo {pages} {106902} (\bibinfo {year} {2024})}\BibitemShut {NoStop}%
\bibitem [{\citenamefont {Jal}\ \emph {et~al.}(2019)\citenamefont {Jal}, \citenamefont {Makita}, \citenamefont {Rösner}, \citenamefont {David}, \citenamefont {Nolting}, \citenamefont {Raabe}, \citenamefont {Savchenko}, \citenamefont {Kleibert}, \citenamefont {Capotondi}, \citenamefont {Pedersoli}, \citenamefont {Raimondi}, \citenamefont {Manfredda}, \citenamefont {Nikolov}, \citenamefont {Liu}, \citenamefont {Merhe}, \citenamefont {Jaouen}, \citenamefont {Gorchon}, \citenamefont {Malinowski}, \citenamefont {Hehn}, \citenamefont {Vodungbo},\ and\ \citenamefont {Lüning}}]{jal_single-shot_2019}%
  \BibitemOpen
  \bibfield  {author} {\bibinfo {author} {\bibfnamefont {E.}~\bibnamefont {Jal}}, \bibinfo {author} {\bibfnamefont {M.}~\bibnamefont {Makita}}, \bibinfo {author} {\bibfnamefont {B.}~\bibnamefont {Rösner}}, \bibinfo {author} {\bibfnamefont {C.}~\bibnamefont {David}}, \bibinfo {author} {\bibfnamefont {F.}~\bibnamefont {Nolting}}, \bibinfo {author} {\bibfnamefont {J.}~\bibnamefont {Raabe}}, \bibinfo {author} {\bibfnamefont {T.}~\bibnamefont {Savchenko}}, \bibinfo {author} {\bibfnamefont {A.}~\bibnamefont {Kleibert}}, \bibinfo {author} {\bibfnamefont {F.}~\bibnamefont {Capotondi}}, \bibinfo {author} {\bibfnamefont {E.}~\bibnamefont {Pedersoli}}, \bibinfo {author} {\bibfnamefont {L.}~\bibnamefont {Raimondi}}, \bibinfo {author} {\bibfnamefont {M.}~\bibnamefont {Manfredda}}, \bibinfo {author} {\bibfnamefont {I.}~\bibnamefont {Nikolov}}, \bibinfo {author} {\bibfnamefont {X.}~\bibnamefont {Liu}}, \bibinfo {author} {\bibfnamefont {A.~e.~d.}\ \bibnamefont {Merhe}}, \bibinfo {author} {\bibfnamefont {N.}~\bibnamefont
  {Jaouen}}, \bibinfo {author} {\bibfnamefont {J.}~\bibnamefont {Gorchon}}, \bibinfo {author} {\bibfnamefont {G.}~\bibnamefont {Malinowski}}, \bibinfo {author} {\bibfnamefont {M.}~\bibnamefont {Hehn}}, \bibinfo {author} {\bibfnamefont {B.}~\bibnamefont {Vodungbo}},\ and\ \bibinfo {author} {\bibfnamefont {J.}~\bibnamefont {Lüning}},\ }\bibfield  {title} {\bibinfo {title} {Single-shot time-resolved magnetic x-ray absorption at a free-electron laser},\ }\href {https://doi.org/10.1103/PhysRevB.99.144305} {\bibfield  {journal} {\bibinfo  {journal} {Physical Review B}\ }\textbf {\bibinfo {volume} {99}},\ \bibinfo {pages} {144305} (\bibinfo {year} {2019})}\BibitemShut {NoStop}%
\bibitem [{\citenamefont {Hennes}\ \emph {et~al.}(2020)\citenamefont {Hennes}, \citenamefont {Rösner}, \citenamefont {Chardonnet}, \citenamefont {Chiuzbaian}, \citenamefont {Delaunay}, \citenamefont {Döring}, \citenamefont {Guzenko}, \citenamefont {Hehn}, \citenamefont {Jarrier}, \citenamefont {Kleibert}, \citenamefont {Lebugle}, \citenamefont {Lüning}, \citenamefont {Malinowski}, \citenamefont {Merhe}, \citenamefont {Naumenko}, \citenamefont {Nikolov}, \citenamefont {Lopez-Quintas}, \citenamefont {Pedersoli}, \citenamefont {Savchenko}, \citenamefont {Watts}, \citenamefont {Zangrando}, \citenamefont {David}, \citenamefont {Capotondi}, \citenamefont {Vodungbo},\ and\ \citenamefont {Jal}}]{hennes_time-resolved_2020}%
  \BibitemOpen
  \bibfield  {author} {\bibinfo {author} {\bibfnamefont {M.}~\bibnamefont {Hennes}}, \bibinfo {author} {\bibfnamefont {B.}~\bibnamefont {Rösner}}, \bibinfo {author} {\bibfnamefont {V.}~\bibnamefont {Chardonnet}}, \bibinfo {author} {\bibfnamefont {G.~S.}\ \bibnamefont {Chiuzbaian}}, \bibinfo {author} {\bibfnamefont {R.}~\bibnamefont {Delaunay}}, \bibinfo {author} {\bibfnamefont {F.}~\bibnamefont {Döring}}, \bibinfo {author} {\bibfnamefont {V.~A.}\ \bibnamefont {Guzenko}}, \bibinfo {author} {\bibfnamefont {M.}~\bibnamefont {Hehn}}, \bibinfo {author} {\bibfnamefont {R.}~\bibnamefont {Jarrier}}, \bibinfo {author} {\bibfnamefont {A.}~\bibnamefont {Kleibert}}, \bibinfo {author} {\bibfnamefont {M.}~\bibnamefont {Lebugle}}, \bibinfo {author} {\bibfnamefont {J.}~\bibnamefont {Lüning}}, \bibinfo {author} {\bibfnamefont {G.}~\bibnamefont {Malinowski}}, \bibinfo {author} {\bibfnamefont {A.}~\bibnamefont {Merhe}}, \bibinfo {author} {\bibfnamefont {D.}~\bibnamefont {Naumenko}}, \bibinfo {author} {\bibfnamefont {I.~P.}\
  \bibnamefont {Nikolov}}, \bibinfo {author} {\bibfnamefont {I.}~\bibnamefont {Lopez-Quintas}}, \bibinfo {author} {\bibfnamefont {E.}~\bibnamefont {Pedersoli}}, \bibinfo {author} {\bibfnamefont {T.}~\bibnamefont {Savchenko}}, \bibinfo {author} {\bibfnamefont {B.}~\bibnamefont {Watts}}, \bibinfo {author} {\bibfnamefont {M.}~\bibnamefont {Zangrando}}, \bibinfo {author} {\bibfnamefont {C.}~\bibnamefont {David}}, \bibinfo {author} {\bibfnamefont {F.}~\bibnamefont {Capotondi}}, \bibinfo {author} {\bibfnamefont {B.}~\bibnamefont {Vodungbo}},\ and\ \bibinfo {author} {\bibfnamefont {E.}~\bibnamefont {Jal}},\ }\bibfield  {title} {\bibinfo {title} {Time-resolved {XUV} absorption spectroscopy and magnetic circular dichroism at the {Ni} {M}$_{2,3}$-edges},\ }\href {https://doi.org/10.3390/app11010325} {\bibfield  {journal} {\bibinfo  {journal} {Applied Sciences}\ }\textbf {\bibinfo {volume} {11}},\ \bibinfo {pages} {325} (\bibinfo {year} {2020})}\BibitemShut {NoStop}%
\bibitem [{\citenamefont {Rösner}\ \emph {et~al.}(2020)\citenamefont {Rösner}, \citenamefont {Vodungbo}, \citenamefont {Chardonnet}, \citenamefont {Döring}, \citenamefont {Guzenko}, \citenamefont {Hennes}, \citenamefont {Kleibert}, \citenamefont {Lebugle}, \citenamefont {Lüning}, \citenamefont {Mahne}, \citenamefont {Merhe}, \citenamefont {Naumenko}, \citenamefont {Nikolov}, \citenamefont {Lopez-Quintas}, \citenamefont {Pedersoli}, \citenamefont {Ribič}, \citenamefont {Savchenko}, \citenamefont {Watts}, \citenamefont {Zangrando}, \citenamefont {Capotondi}, \citenamefont {David},\ and\ \citenamefont {Jal}}]{rosner_simultaneous_2020}%
  \BibitemOpen
  \bibfield  {author} {\bibinfo {author} {\bibfnamefont {B.}~\bibnamefont {Rösner}}, \bibinfo {author} {\bibfnamefont {B.}~\bibnamefont {Vodungbo}}, \bibinfo {author} {\bibfnamefont {V.}~\bibnamefont {Chardonnet}}, \bibinfo {author} {\bibfnamefont {F.}~\bibnamefont {Döring}}, \bibinfo {author} {\bibfnamefont {V.~A.}\ \bibnamefont {Guzenko}}, \bibinfo {author} {\bibfnamefont {M.}~\bibnamefont {Hennes}}, \bibinfo {author} {\bibfnamefont {A.}~\bibnamefont {Kleibert}}, \bibinfo {author} {\bibfnamefont {M.}~\bibnamefont {Lebugle}}, \bibinfo {author} {\bibfnamefont {J.}~\bibnamefont {Lüning}}, \bibinfo {author} {\bibfnamefont {N.}~\bibnamefont {Mahne}}, \bibinfo {author} {\bibfnamefont {A.}~\bibnamefont {Merhe}}, \bibinfo {author} {\bibfnamefont {D.}~\bibnamefont {Naumenko}}, \bibinfo {author} {\bibfnamefont {I.~P.}\ \bibnamefont {Nikolov}}, \bibinfo {author} {\bibfnamefont {I.}~\bibnamefont {Lopez-Quintas}}, \bibinfo {author} {\bibfnamefont {E.}~\bibnamefont {Pedersoli}}, \bibinfo {author} {\bibfnamefont
  {P.~R.}\ \bibnamefont {Ribič}}, \bibinfo {author} {\bibfnamefont {T.}~\bibnamefont {Savchenko}}, \bibinfo {author} {\bibfnamefont {B.}~\bibnamefont {Watts}}, \bibinfo {author} {\bibfnamefont {M.}~\bibnamefont {Zangrando}}, \bibinfo {author} {\bibfnamefont {F.}~\bibnamefont {Capotondi}}, \bibinfo {author} {\bibfnamefont {C.}~\bibnamefont {David}},\ and\ \bibinfo {author} {\bibfnamefont {E.}~\bibnamefont {Jal}},\ }\bibfield  {title} {\bibinfo {title} {Simultaneous two-color snapshot view on ultrafast charge and spin dynamics in a {Fe}-{Cu}-{Ni} tri-layer},\ }\href {https://doi.org/10.1063/4.0000033} {\bibfield  {journal} {\bibinfo  {journal} {Structural Dynamics}\ }\textbf {\bibinfo {volume} {7}},\ \bibinfo {pages} {054302} (\bibinfo {year} {2020})}\BibitemShut {NoStop}%
\bibitem [{\citenamefont {Kfir}\ \emph {et~al.}(2014)\citenamefont {Kfir}, \citenamefont {Grychtol}, \citenamefont {Turgut}, \citenamefont {Knut}, \citenamefont {Zusin}, \citenamefont {Popmintchev}, \citenamefont {Popmintchev}, \citenamefont {Nembach}, \citenamefont {Shaw}, \citenamefont {Fleischer}, \citenamefont {Kapteyn}, \citenamefont {Murnane},\ and\ \citenamefont {Cohen}}]{Kfir2014}%
  \BibitemOpen
  \bibfield  {author} {\bibinfo {author} {\bibfnamefont {O.}~\bibnamefont {Kfir}}, \bibinfo {author} {\bibfnamefont {P.}~\bibnamefont {Grychtol}}, \bibinfo {author} {\bibfnamefont {E.}~\bibnamefont {Turgut}}, \bibinfo {author} {\bibfnamefont {R.}~\bibnamefont {Knut}}, \bibinfo {author} {\bibfnamefont {D.}~\bibnamefont {Zusin}}, \bibinfo {author} {\bibfnamefont {D.}~\bibnamefont {Popmintchev}}, \bibinfo {author} {\bibfnamefont {T.}~\bibnamefont {Popmintchev}}, \bibinfo {author} {\bibfnamefont {H.}~\bibnamefont {Nembach}}, \bibinfo {author} {\bibfnamefont {J.~M.}\ \bibnamefont {Shaw}}, \bibinfo {author} {\bibfnamefont {A.}~\bibnamefont {Fleischer}}, \bibinfo {author} {\bibfnamefont {H.}~\bibnamefont {Kapteyn}}, \bibinfo {author} {\bibfnamefont {M.}~\bibnamefont {Murnane}},\ and\ \bibinfo {author} {\bibfnamefont {O.}~\bibnamefont {Cohen}},\ }\bibfield  {title} {\bibinfo {title} {Generation of bright phase-matched circularly-polarized extreme ultraviolet high harmonics},\ }\href
  {https://doi.org/10.1038/nphoton.2014.293} {\bibfield  {journal} {\bibinfo  {journal} {Nature Photonics}\ }\textbf {\bibinfo {volume} {9}},\ \bibinfo {pages} {99} (\bibinfo {year} {2014})}\BibitemShut {NoStop}%
\bibitem [{\citenamefont {Fan}\ \emph {et~al.}(2015)\citenamefont {Fan}, \citenamefont {Grychtol}, \citenamefont {Knut}, \citenamefont {Hernández-García}, \citenamefont {Hickstein}, \citenamefont {Zusin}, \citenamefont {Gentry}, \citenamefont {Dollar}, \citenamefont {Mancuso}, \citenamefont {Hogle}, \citenamefont {Kfir}, \citenamefont {Legut}, \citenamefont {Carva}, \citenamefont {Ellis}, \citenamefont {Dorney}, \citenamefont {Chen}, \citenamefont {Shpyrko}, \citenamefont {Fullerton}, \citenamefont {Cohen}, \citenamefont {Oppeneer}, \citenamefont {Milošević}, \citenamefont {Becker}, \citenamefont {Jaroń-Becker}, \citenamefont {Popmintchev}, \citenamefont {Murnane},\ and\ \citenamefont {Kapteyn}}]{Fan2015}%
  \BibitemOpen
  \bibfield  {author} {\bibinfo {author} {\bibfnamefont {T.}~\bibnamefont {Fan}}, \bibinfo {author} {\bibfnamefont {P.}~\bibnamefont {Grychtol}}, \bibinfo {author} {\bibfnamefont {R.}~\bibnamefont {Knut}}, \bibinfo {author} {\bibfnamefont {C.}~\bibnamefont {Hernández-García}}, \bibinfo {author} {\bibfnamefont {D.~D.}\ \bibnamefont {Hickstein}}, \bibinfo {author} {\bibfnamefont {D.}~\bibnamefont {Zusin}}, \bibinfo {author} {\bibfnamefont {C.}~\bibnamefont {Gentry}}, \bibinfo {author} {\bibfnamefont {F.~J.}\ \bibnamefont {Dollar}}, \bibinfo {author} {\bibfnamefont {C.~A.}\ \bibnamefont {Mancuso}}, \bibinfo {author} {\bibfnamefont {C.~W.}\ \bibnamefont {Hogle}}, \bibinfo {author} {\bibfnamefont {O.}~\bibnamefont {Kfir}}, \bibinfo {author} {\bibfnamefont {D.}~\bibnamefont {Legut}}, \bibinfo {author} {\bibfnamefont {K.}~\bibnamefont {Carva}}, \bibinfo {author} {\bibfnamefont {J.~L.}\ \bibnamefont {Ellis}}, \bibinfo {author} {\bibfnamefont {K.~M.}\ \bibnamefont {Dorney}}, \bibinfo {author} {\bibfnamefont
  {C.}~\bibnamefont {Chen}}, \bibinfo {author} {\bibfnamefont {O.~G.}\ \bibnamefont {Shpyrko}}, \bibinfo {author} {\bibfnamefont {E.~E.}\ \bibnamefont {Fullerton}}, \bibinfo {author} {\bibfnamefont {O.}~\bibnamefont {Cohen}}, \bibinfo {author} {\bibfnamefont {P.~M.}\ \bibnamefont {Oppeneer}}, \bibinfo {author} {\bibfnamefont {D.~B.}\ \bibnamefont {Milošević}}, \bibinfo {author} {\bibfnamefont {A.}~\bibnamefont {Becker}}, \bibinfo {author} {\bibfnamefont {A.~A.}\ \bibnamefont {Jaroń-Becker}}, \bibinfo {author} {\bibfnamefont {T.}~\bibnamefont {Popmintchev}}, \bibinfo {author} {\bibfnamefont {M.~M.}\ \bibnamefont {Murnane}},\ and\ \bibinfo {author} {\bibfnamefont {H.~C.}\ \bibnamefont {Kapteyn}},\ }\bibfield  {title} {\bibinfo {title} {Bright circularly polarized soft x-ray high harmonics for x-ray magnetic circular dichroism},\ }\href {https://doi.org/10.1073/pnas.1519666112} {\bibfield  {journal} {\bibinfo  {journal} {Proceedings of the National Academy of Sciences}\ }\textbf {\bibinfo {volume} {112}},\
  \bibinfo {pages} {14206} (\bibinfo {year} {2015})}\BibitemShut {NoStop}%
\bibitem [{\citenamefont {Lambert}\ \emph {et~al.}(2015)\citenamefont {Lambert}, \citenamefont {Vodungbo}, \citenamefont {Gautier}, \citenamefont {Mahieu}, \citenamefont {Malka}, \citenamefont {Sebban}, \citenamefont {Zeitoun}, \citenamefont {Luning}, \citenamefont {Perron}, \citenamefont {Andreev}, \citenamefont {Stremoukhov}, \citenamefont {Ardana-Lamas}, \citenamefont {Dax}, \citenamefont {Hauri}, \citenamefont {Sardinha},\ and\ \citenamefont {Fajardo}}]{Lambert2015}%
  \BibitemOpen
  \bibfield  {author} {\bibinfo {author} {\bibfnamefont {G.}~\bibnamefont {Lambert}}, \bibinfo {author} {\bibfnamefont {B.}~\bibnamefont {Vodungbo}}, \bibinfo {author} {\bibfnamefont {J.}~\bibnamefont {Gautier}}, \bibinfo {author} {\bibfnamefont {B.}~\bibnamefont {Mahieu}}, \bibinfo {author} {\bibfnamefont {V.}~\bibnamefont {Malka}}, \bibinfo {author} {\bibfnamefont {S.}~\bibnamefont {Sebban}}, \bibinfo {author} {\bibfnamefont {P.}~\bibnamefont {Zeitoun}}, \bibinfo {author} {\bibfnamefont {J.}~\bibnamefont {Luning}}, \bibinfo {author} {\bibfnamefont {J.}~\bibnamefont {Perron}}, \bibinfo {author} {\bibfnamefont {A.}~\bibnamefont {Andreev}}, \bibinfo {author} {\bibfnamefont {S.}~\bibnamefont {Stremoukhov}}, \bibinfo {author} {\bibfnamefont {F.}~\bibnamefont {Ardana-Lamas}}, \bibinfo {author} {\bibfnamefont {A.}~\bibnamefont {Dax}}, \bibinfo {author} {\bibfnamefont {C.~P.}\ \bibnamefont {Hauri}}, \bibinfo {author} {\bibfnamefont {A.}~\bibnamefont {Sardinha}},\ and\ \bibinfo {author} {\bibfnamefont
  {M.}~\bibnamefont {Fajardo}},\ }\bibfield  {title} {\bibinfo {title} {Towards enabling femtosecond helicity-dependent spectroscopy with high-harmonic sources},\ }\href {https://www.nature.com/articles/ncomms7167} {\bibfield  {journal} {\bibinfo  {journal} {Nature Communications}\ }\textbf {\bibinfo {volume} {6}} (\bibinfo {year} {2015})}\BibitemShut {NoStop}%
\bibitem [{\citenamefont {Vodungbo}\ \emph {et~al.}(2011)\citenamefont {Vodungbo}, \citenamefont {Barszczak~Sardinha}, \citenamefont {Gautier}, \citenamefont {Lambert}, \citenamefont {Valentin}, \citenamefont {Lozano}, \citenamefont {Iaquaniello}, \citenamefont {Delmotte}, \citenamefont {Sebban}, \citenamefont {Lüning},\ and\ \citenamefont {Zeitoun}}]{Vodungbo2011}%
  \BibitemOpen
  \bibfield  {author} {\bibinfo {author} {\bibfnamefont {B.}~\bibnamefont {Vodungbo}}, \bibinfo {author} {\bibfnamefont {A.}~\bibnamefont {Barszczak~Sardinha}}, \bibinfo {author} {\bibfnamefont {J.}~\bibnamefont {Gautier}}, \bibinfo {author} {\bibfnamefont {G.}~\bibnamefont {Lambert}}, \bibinfo {author} {\bibfnamefont {C.}~\bibnamefont {Valentin}}, \bibinfo {author} {\bibfnamefont {M.}~\bibnamefont {Lozano}}, \bibinfo {author} {\bibfnamefont {G.}~\bibnamefont {Iaquaniello}}, \bibinfo {author} {\bibfnamefont {F.}~\bibnamefont {Delmotte}}, \bibinfo {author} {\bibfnamefont {S.}~\bibnamefont {Sebban}}, \bibinfo {author} {\bibfnamefont {J.}~\bibnamefont {Lüning}},\ and\ \bibinfo {author} {\bibfnamefont {P.}~\bibnamefont {Zeitoun}},\ }\bibfield  {title} {\bibinfo {title} {Polarization control of high order harmonics in the {EUV} photon energy range},\ }\href {https://doi.org/10.1364/oe.19.004346} {\bibfield  {journal} {\bibinfo  {journal} {Optics Express}\ }\textbf {\bibinfo {volume} {19}},\ \bibinfo {pages}
  {4346} (\bibinfo {year} {2011})}\BibitemShut {NoStop}%
\bibitem [{\citenamefont {von Korff~Schmising}\ \emph {et~al.}(2017)\citenamefont {von Korff~Schmising}, \citenamefont {Weder}, \citenamefont {Noll}, \citenamefont {Pfau}, \citenamefont {Hennecke}, \citenamefont {Strüber}, \citenamefont {Radu}, \citenamefont {Schneider}, \citenamefont {Staeck}, \citenamefont {Günther}, \citenamefont {Lüning}, \citenamefont {Merhe}, \citenamefont {Buck}, \citenamefont {Hartmann}, \citenamefont {Viefhaus}, \citenamefont {Treusch},\ and\ \citenamefont {Eisebitt}}]{KorffSchmising2017}%
  \BibitemOpen
  \bibfield  {author} {\bibinfo {author} {\bibfnamefont {C.}~\bibnamefont {von Korff~Schmising}}, \bibinfo {author} {\bibfnamefont {D.}~\bibnamefont {Weder}}, \bibinfo {author} {\bibfnamefont {T.}~\bibnamefont {Noll}}, \bibinfo {author} {\bibfnamefont {B.}~\bibnamefont {Pfau}}, \bibinfo {author} {\bibfnamefont {M.}~\bibnamefont {Hennecke}}, \bibinfo {author} {\bibfnamefont {C.}~\bibnamefont {Strüber}}, \bibinfo {author} {\bibfnamefont {I.}~\bibnamefont {Radu}}, \bibinfo {author} {\bibfnamefont {M.}~\bibnamefont {Schneider}}, \bibinfo {author} {\bibfnamefont {S.}~\bibnamefont {Staeck}}, \bibinfo {author} {\bibfnamefont {C.~M.}\ \bibnamefont {Günther}}, \bibinfo {author} {\bibfnamefont {J.}~\bibnamefont {Lüning}}, \bibinfo {author} {\bibfnamefont {A.~e.~d.}\ \bibnamefont {Merhe}}, \bibinfo {author} {\bibfnamefont {J.}~\bibnamefont {Buck}}, \bibinfo {author} {\bibfnamefont {G.}~\bibnamefont {Hartmann}}, \bibinfo {author} {\bibfnamefont {J.}~\bibnamefont {Viefhaus}}, \bibinfo {author} {\bibfnamefont
  {R.}~\bibnamefont {Treusch}},\ and\ \bibinfo {author} {\bibfnamefont {S.}~\bibnamefont {Eisebitt}},\ }\bibfield  {title} {\bibinfo {title} {Generating circularly polarized radiation in the extreme ultraviolet spectral range at the free-electron laser {FLASH}},\ }\href {https://pubs.aip.org/aip/rsi/article/88/5/053903/1003209/Generating-circularly-polarized-radiation-in-the} {\bibfield  {journal} {\bibinfo  {journal} {Review of Scientific Instruments}\ }\textbf {\bibinfo {volume} {88}} (\bibinfo {year} {2017})}\BibitemShut {NoStop}%
\bibitem [{\citenamefont {J~Lopes}\ \emph {et~al.}(2021)\citenamefont {J~Lopes}, \citenamefont {Czubak}, \citenamefont {Zallo}, \citenamefont {Figueroa}, \citenamefont {Guillemard}, \citenamefont {Valvidares}, \citenamefont {Rubio-Zuazo}, \citenamefont {López-Sanchéz}, \citenamefont {Valenzuela}, \citenamefont {Hanke},\ and\ \citenamefont {Ramsteiner}}]{lopes_large-area_2021}%
  \BibitemOpen
  \bibfield  {author} {\bibinfo {author} {\bibfnamefont {J.~M.}\ \bibnamefont {J~Lopes}}, \bibinfo {author} {\bibfnamefont {D.}~\bibnamefont {Czubak}}, \bibinfo {author} {\bibfnamefont {E.}~\bibnamefont {Zallo}}, \bibinfo {author} {\bibfnamefont {A.~I.}\ \bibnamefont {Figueroa}}, \bibinfo {author} {\bibfnamefont {C.}~\bibnamefont {Guillemard}}, \bibinfo {author} {\bibfnamefont {M.}~\bibnamefont {Valvidares}}, \bibinfo {author} {\bibfnamefont {J.}~\bibnamefont {Rubio-Zuazo}}, \bibinfo {author} {\bibfnamefont {J.}~\bibnamefont {López-Sanchéz}}, \bibinfo {author} {\bibfnamefont {S.~O.}\ \bibnamefont {Valenzuela}}, \bibinfo {author} {\bibfnamefont {M.}~\bibnamefont {Hanke}},\ and\ \bibinfo {author} {\bibfnamefont {M.}~\bibnamefont {Ramsteiner}},\ }\bibfield  {title} {\bibinfo {title} {Large-area van der {Waals} epitaxy and magnetic characterization of {Fe}$_{\textrm{3}}$ {GeTe}$_{\textrm{2}}$ films on graphene},\ }\href {https://doi.org/10.1088/2053-1583/ac171d} {\bibfield  {journal} {\bibinfo  {journal} {2D
  Materials}\ }\textbf {\bibinfo {volume} {8}},\ \bibinfo {pages} {041001} (\bibinfo {year} {2021})}\BibitemShut {NoStop}%
\bibitem [{\citenamefont {Oppeneer}\ \emph {et~al.}()\citenamefont {Oppeneer}, \citenamefont {Mertins},\ and\ \citenamefont {Zaharko}}]{oppeneer_2003}%
  \BibitemOpen
  \bibfield  {author} {\bibinfo {author} {\bibfnamefont {P.~M.}\ \bibnamefont {Oppeneer}}, \bibinfo {author} {\bibfnamefont {H.-C.}\ \bibnamefont {Mertins}},\ and\ \bibinfo {author} {\bibfnamefont {O.}~\bibnamefont {Zaharko}},\ }\bibfield  {title} {\bibinfo {title} {Alternative geometries for the determination of x-ray magneto-optical coefficients},\ }\href {https://doi.org/10.1088/0953-8984/15/45/018} {\bibfield  {journal} {\bibinfo  {journal} {Journal of Physics: Condensed Matter}\ }\textbf {\bibinfo {volume} {15}},\ \bibinfo {pages} {7803}}\BibitemShut {NoStop}%
\bibitem [{\citenamefont {Yao}\ \emph {et~al.}(2020{\natexlab{b}})\citenamefont {Yao}, \citenamefont {Willems}, \citenamefont {von Korff~Schmising}, \citenamefont {Strüber}, \citenamefont {Hessing}, \citenamefont {Pfau}, \citenamefont {Schick}, \citenamefont {Engel}, \citenamefont {Gerlinger}, \citenamefont {Schneider},\ and\ \citenamefont {Eisebitt}}]{yao_tabletop_2020}%
  \BibitemOpen
  \bibfield  {author} {\bibinfo {author} {\bibfnamefont {K.}~\bibnamefont {Yao}}, \bibinfo {author} {\bibfnamefont {F.}~\bibnamefont {Willems}}, \bibinfo {author} {\bibfnamefont {C.}~\bibnamefont {von Korff~Schmising}}, \bibinfo {author} {\bibfnamefont {C.}~\bibnamefont {Strüber}}, \bibinfo {author} {\bibfnamefont {P.}~\bibnamefont {Hessing}}, \bibinfo {author} {\bibfnamefont {B.}~\bibnamefont {Pfau}}, \bibinfo {author} {\bibfnamefont {D.}~\bibnamefont {Schick}}, \bibinfo {author} {\bibfnamefont {D.}~\bibnamefont {Engel}}, \bibinfo {author} {\bibfnamefont {K.}~\bibnamefont {Gerlinger}}, \bibinfo {author} {\bibfnamefont {M.}~\bibnamefont {Schneider}},\ and\ \bibinfo {author} {\bibfnamefont {S.}~\bibnamefont {Eisebitt}},\ }\bibfield  {title} {\bibinfo {title} {A tabletop setup for ultrafast helicity-dependent and element-specific absorption spectroscopy and scattering in the extreme ultraviolet spectral range},\ }\href {https://doi.org/10.1063/5.0013928} {\bibfield  {journal} {\bibinfo  {journal} {Review of
  Scientific Instruments}\ }\textbf {\bibinfo {volume} {91}},\ \bibinfo {pages} {093001} (\bibinfo {year} {2020}{\natexlab{b}})}\BibitemShut {NoStop}%
\bibitem [{\citenamefont {Schick}(2021)}]{schick_udkm1dsim_2021}%
  \BibitemOpen
  \bibfield  {author} {\bibinfo {author} {\bibfnamefont {D.}~\bibnamefont {Schick}},\ }\bibfield  {title} {\bibinfo {title} {{udkm1Dsim} – a {Python} toolbox for simulating {1D} ultrafast dynamics in condensed matter},\ }\href {https://doi.org/10.1016/j.cpc.2021.108031} {\bibfield  {journal} {\bibinfo  {journal} {Computer Physics Communications}\ }\textbf {\bibinfo {volume} {266}},\ \bibinfo {pages} {108031} (\bibinfo {year} {2021})}\BibitemShut {NoStop}%
\bibitem [{\citenamefont {Elzo}\ \emph {et~al.}(2012)\citenamefont {Elzo}, \citenamefont {Jal}, \citenamefont {Bunau}, \citenamefont {Grenier}, \citenamefont {Joly}, \citenamefont {Ramos}, \citenamefont {Tolentino}, \citenamefont {Tonnerre},\ and\ \citenamefont {Jaouen}}]{Elzo2012}%
  \BibitemOpen
  \bibfield  {author} {\bibinfo {author} {\bibfnamefont {M.}~\bibnamefont {Elzo}}, \bibinfo {author} {\bibfnamefont {E.}~\bibnamefont {Jal}}, \bibinfo {author} {\bibfnamefont {O.}~\bibnamefont {Bunau}}, \bibinfo {author} {\bibfnamefont {S.}~\bibnamefont {Grenier}}, \bibinfo {author} {\bibfnamefont {Y.}~\bibnamefont {Joly}}, \bibinfo {author} {\bibfnamefont {A.}~\bibnamefont {Ramos}}, \bibinfo {author} {\bibfnamefont {H.}~\bibnamefont {Tolentino}}, \bibinfo {author} {\bibfnamefont {J.}~\bibnamefont {Tonnerre}},\ and\ \bibinfo {author} {\bibfnamefont {N.}~\bibnamefont {Jaouen}},\ }\bibfield  {title} {\bibinfo {title} {X-ray resonant magnetic reflectivity of stratified magnetic structures: Eigenwave formalism and application to a w/fe/w trilayer},\ }\href {https://doi.org/10.1016/j.jmmm.2011.07.019} {\bibfield  {journal} {\bibinfo  {journal} {Journal of Magnetism and Magnetic Materials}\ }\textbf {\bibinfo {volume} {324}},\ \bibinfo {pages} {105} (\bibinfo {year} {2012})}\BibitemShut {NoStop}%
\bibitem [{\citenamefont {Henke}\ \emph {et~al.}(1993)\citenamefont {Henke}, \citenamefont {Gullikson},\ and\ \citenamefont {Davis}}]{Henke1993}%
  \BibitemOpen
  \bibfield  {author} {\bibinfo {author} {\bibfnamefont {B.}~\bibnamefont {Henke}}, \bibinfo {author} {\bibfnamefont {E.}~\bibnamefont {Gullikson}},\ and\ \bibinfo {author} {\bibfnamefont {J.}~\bibnamefont {Davis}},\ }\bibfield  {title} {\bibinfo {title} {X-ray interactions: Photoabsorption, scattering, transmission, and reflection at {E} = 50-30,000 e{V}, {Z} = 1-92},\ }\href {https://doi.org/https://doi.org/10.1006/adnd.1993.1013} {\bibfield  {journal} {\bibinfo  {journal} {Atomic Data and Nuclear Data Tables}\ }\textbf {\bibinfo {volume} {54}},\ \bibinfo {pages} {181} (\bibinfo {year} {1993})}\BibitemShut {NoStop}%
\bibitem [{\citenamefont {Willems}\ \emph {et~al.}(2019)\citenamefont {Willems}, \citenamefont {Sharma}, \citenamefont {{v. Korff Schmising}}, \citenamefont {Dewhurst}, \citenamefont {Salemi}, \citenamefont {Schick}, \citenamefont {Hessing}, \citenamefont {Str{\"{u}}ber}, \citenamefont {Engel},\ and\ \citenamefont {Eisebitt}}]{Willems2019}%
  \BibitemOpen
  \bibfield  {author} {\bibinfo {author} {\bibfnamefont {F.}~\bibnamefont {Willems}}, \bibinfo {author} {\bibfnamefont {S.}~\bibnamefont {Sharma}}, \bibinfo {author} {\bibfnamefont {C.}~\bibnamefont {{v. Korff Schmising}}}, \bibinfo {author} {\bibfnamefont {J.~K.}\ \bibnamefont {Dewhurst}}, \bibinfo {author} {\bibfnamefont {L.}~\bibnamefont {Salemi}}, \bibinfo {author} {\bibfnamefont {D.}~\bibnamefont {Schick}}, \bibinfo {author} {\bibfnamefont {P.}~\bibnamefont {Hessing}}, \bibinfo {author} {\bibfnamefont {C.}~\bibnamefont {Str{\"{u}}ber}}, \bibinfo {author} {\bibfnamefont {W.~D.}\ \bibnamefont {Engel}},\ and\ \bibinfo {author} {\bibfnamefont {S.}~\bibnamefont {Eisebitt}},\ }\bibfield  {title} {\bibinfo {title} {{Magneto-Optical Functions at the 3p Resonances of Fe, Co, and Ni: Ab initio Description and Experiment}},\ }\href {https://doi.org/10.1103/PhysRevLett.122.217202} {\bibfield  {journal} {\bibinfo  {journal} {Phys. Rev. Lett.}\ }\textbf {\bibinfo {volume} {122}},\ \bibinfo {pages} {217202} (\bibinfo
  {year} {2019})}\BibitemShut {NoStop}%
\bibitem [{\citenamefont {Booklet}(2001)}]{booklet2001x}%
  \BibitemOpen
  \bibfield  {author} {\bibinfo {author} {\bibfnamefont {X.-R.~D.}\ \bibnamefont {Booklet}},\ }\bibfield  {title} {\bibinfo {title} {X-ray data booklet},\ }\href@noop {} {\bibfield  {journal} {\bibinfo  {journal} {Laboratory, Univ. California}\ } (\bibinfo {year} {2001})}\BibitemShut {NoStop}%
\bibitem [{\citenamefont {Steinbach}\ \emph {et~al.}(2021)\citenamefont {Steinbach}, \citenamefont {Schick}, \citenamefont {Von Korff~Schmising}, \citenamefont {Yao}, \citenamefont {Borchert}, \citenamefont {Engel},\ and\ \citenamefont {Eisebitt}}]{steinbach_wide-field_2021}%
  \BibitemOpen
  \bibfield  {author} {\bibinfo {author} {\bibfnamefont {F.}~\bibnamefont {Steinbach}}, \bibinfo {author} {\bibfnamefont {D.}~\bibnamefont {Schick}}, \bibinfo {author} {\bibfnamefont {C.}~\bibnamefont {Von Korff~Schmising}}, \bibinfo {author} {\bibfnamefont {K.}~\bibnamefont {Yao}}, \bibinfo {author} {\bibfnamefont {M.}~\bibnamefont {Borchert}}, \bibinfo {author} {\bibfnamefont {W.~D.}\ \bibnamefont {Engel}},\ and\ \bibinfo {author} {\bibfnamefont {S.}~\bibnamefont {Eisebitt}},\ }\bibfield  {title} {\bibinfo {title} {Wide-field magneto-optical microscope to access quantitative magnetization dynamics with femtosecond temporal and sub-micrometer spatial resolution},\ }\href {https://doi.org/10.1063/5.0060091} {\bibfield  {journal} {\bibinfo  {journal} {Journal of Applied Physics}\ }\textbf {\bibinfo {volume} {130}},\ \bibinfo {pages} {083905} (\bibinfo {year} {2021})}\BibitemShut {NoStop}%
\bibitem [{\citenamefont {Hashimoto}\ \emph {et~al.}(2014)\citenamefont {Hashimoto}, \citenamefont {Khorsand}, \citenamefont {Savoini}, \citenamefont {Koene}, \citenamefont {Bossini}, \citenamefont {Tsukamoto}, \citenamefont {Itoh}, \citenamefont {Ohtsuka}, \citenamefont {Aoshima}, \citenamefont {Kimel}, \citenamefont {Kirilyuk},\ and\ \citenamefont {Rasing}}]{hashimoto_ultrafast_2014}%
  \BibitemOpen
  \bibfield  {author} {\bibinfo {author} {\bibfnamefont {Y.}~\bibnamefont {Hashimoto}}, \bibinfo {author} {\bibfnamefont {A.~R.}\ \bibnamefont {Khorsand}}, \bibinfo {author} {\bibfnamefont {M.}~\bibnamefont {Savoini}}, \bibinfo {author} {\bibfnamefont {B.}~\bibnamefont {Koene}}, \bibinfo {author} {\bibfnamefont {D.}~\bibnamefont {Bossini}}, \bibinfo {author} {\bibfnamefont {A.}~\bibnamefont {Tsukamoto}}, \bibinfo {author} {\bibfnamefont {A.}~\bibnamefont {Itoh}}, \bibinfo {author} {\bibfnamefont {Y.}~\bibnamefont {Ohtsuka}}, \bibinfo {author} {\bibfnamefont {K.}~\bibnamefont {Aoshima}}, \bibinfo {author} {\bibfnamefont {A.~V.}\ \bibnamefont {Kimel}}, \bibinfo {author} {\bibfnamefont {A.}~\bibnamefont {Kirilyuk}},\ and\ \bibinfo {author} {\bibfnamefont {T.}~\bibnamefont {Rasing}},\ }\bibfield  {title} {\bibinfo {title} {Ultrafast time-resolved magneto-optical imaging of all-optical switching in {GdFeCo} with femtosecond time-resolution and a $\mu$m spatial-resolution},\ }\href
  {https://doi.org/10.1063/1.4880015} {\bibfield  {journal} {\bibinfo  {journal} {Review of Scientific Instruments}\ }\textbf {\bibinfo {volume} {85}},\ \bibinfo {pages} {063702} (\bibinfo {year} {2014})}\BibitemShut {NoStop}%
\bibitem [{\citenamefont {Jana}\ \emph {et~al.}(2017)\citenamefont {Jana}, \citenamefont {Terschlüsen}, \citenamefont {Stefanuik}, \citenamefont {Plogmaker}, \citenamefont {Troisi}, \citenamefont {Malik}, \citenamefont {Svanqvist}, \citenamefont {Knut}, \citenamefont {Söderström},\ and\ \citenamefont {Karis}}]{jana_setup_2017}%
  \BibitemOpen
  \bibfield  {author} {\bibinfo {author} {\bibfnamefont {S.}~\bibnamefont {Jana}}, \bibinfo {author} {\bibfnamefont {J.~A.}\ \bibnamefont {Terschlüsen}}, \bibinfo {author} {\bibfnamefont {R.}~\bibnamefont {Stefanuik}}, \bibinfo {author} {\bibfnamefont {S.}~\bibnamefont {Plogmaker}}, \bibinfo {author} {\bibfnamefont {S.}~\bibnamefont {Troisi}}, \bibinfo {author} {\bibfnamefont {R.~S.}\ \bibnamefont {Malik}}, \bibinfo {author} {\bibfnamefont {M.}~\bibnamefont {Svanqvist}}, \bibinfo {author} {\bibfnamefont {R.}~\bibnamefont {Knut}}, \bibinfo {author} {\bibfnamefont {J.}~\bibnamefont {Söderström}},\ and\ \bibinfo {author} {\bibfnamefont {O.}~\bibnamefont {Karis}},\ }\bibfield  {title} {\bibinfo {title} {A setup for element specific magnetization dynamics using the transverse magneto-optic {Kerr} effect in the energy range of 30-72 {eV}},\ }\href {https://doi.org/10.1063/1.4978907} {\bibfield  {journal} {\bibinfo  {journal} {Review of Scientific Instruments}\ }\textbf {\bibinfo {volume} {88}},\ \bibinfo {pages}
  {033113} (\bibinfo {year} {2017})}\BibitemShut {NoStop}%
\bibitem [{\citenamefont {Jana}\ \emph {et~al.}(2022)\citenamefont {Jana}, \citenamefont {Knut}, \citenamefont {Muralidhar}, \citenamefont {Malik}, \citenamefont {Stefanuik}, \citenamefont {Åkerman}, \citenamefont {Karis}, \citenamefont {Schüßler-Langeheine},\ and\ \citenamefont {Pontius}}]{jana_experimental_2022}%
  \BibitemOpen
  \bibfield  {author} {\bibinfo {author} {\bibfnamefont {S.}~\bibnamefont {Jana}}, \bibinfo {author} {\bibfnamefont {R.}~\bibnamefont {Knut}}, \bibinfo {author} {\bibfnamefont {S.}~\bibnamefont {Muralidhar}}, \bibinfo {author} {\bibfnamefont {R.~S.}\ \bibnamefont {Malik}}, \bibinfo {author} {\bibfnamefont {R.}~\bibnamefont {Stefanuik}}, \bibinfo {author} {\bibfnamefont {J.}~\bibnamefont {Åkerman}}, \bibinfo {author} {\bibfnamefont {O.}~\bibnamefont {Karis}}, \bibinfo {author} {\bibfnamefont {C.}~\bibnamefont {Schüßler-Langeheine}},\ and\ \bibinfo {author} {\bibfnamefont {N.}~\bibnamefont {Pontius}},\ }\bibfield  {title} {\bibinfo {title} {Experimental confirmation of the delayed {Ni} demagnetization in {FeNi} alloy},\ }\href {https://doi.org/10.1063/5.0080331} {\bibfield  {journal} {\bibinfo  {journal} {Applied Physics Letters}\ }\textbf {\bibinfo {volume} {120}},\ \bibinfo {pages} {102404} (\bibinfo {year} {2022})}\BibitemShut {NoStop}%
\bibitem [{\citenamefont {Hofherr}\ \emph {et~al.}(2020)\citenamefont {Hofherr}, \citenamefont {Häuser}, \citenamefont {Dewhurst}, \citenamefont {Tengdin}, \citenamefont {Sakshath}, \citenamefont {Nembach}, \citenamefont {Weber}, \citenamefont {Shaw}, \citenamefont {Silva}, \citenamefont {Kapteyn}, \citenamefont {Cinchetti}, \citenamefont {Rethfeld}, \citenamefont {Murnane}, \citenamefont {Steil}, \citenamefont {Stadtmüller}, \citenamefont {Sharma}, \citenamefont {Aeschlimann},\ and\ \citenamefont {Mathias}}]{hofherr_ultrafast_2020}%
  \BibitemOpen
  \bibfield  {author} {\bibinfo {author} {\bibfnamefont {M.}~\bibnamefont {Hofherr}}, \bibinfo {author} {\bibfnamefont {S.}~\bibnamefont {Häuser}}, \bibinfo {author} {\bibfnamefont {J.~K.}\ \bibnamefont {Dewhurst}}, \bibinfo {author} {\bibfnamefont {P.}~\bibnamefont {Tengdin}}, \bibinfo {author} {\bibfnamefont {S.}~\bibnamefont {Sakshath}}, \bibinfo {author} {\bibfnamefont {H.~T.}\ \bibnamefont {Nembach}}, \bibinfo {author} {\bibfnamefont {S.~T.}\ \bibnamefont {Weber}}, \bibinfo {author} {\bibfnamefont {J.~M.}\ \bibnamefont {Shaw}}, \bibinfo {author} {\bibfnamefont {T.~J.}\ \bibnamefont {Silva}}, \bibinfo {author} {\bibfnamefont {H.~C.}\ \bibnamefont {Kapteyn}}, \bibinfo {author} {\bibfnamefont {M.}~\bibnamefont {Cinchetti}}, \bibinfo {author} {\bibfnamefont {B.}~\bibnamefont {Rethfeld}}, \bibinfo {author} {\bibfnamefont {M.~M.}\ \bibnamefont {Murnane}}, \bibinfo {author} {\bibfnamefont {D.}~\bibnamefont {Steil}}, \bibinfo {author} {\bibfnamefont {B.}~\bibnamefont {Stadtmüller}}, \bibinfo {author}
  {\bibfnamefont {S.}~\bibnamefont {Sharma}}, \bibinfo {author} {\bibfnamefont {M.}~\bibnamefont {Aeschlimann}},\ and\ \bibinfo {author} {\bibfnamefont {S.}~\bibnamefont {Mathias}},\ }\bibfield  {title} {\bibinfo {title} {Ultrafast optically induced spin transfer in ferromagnetic alloys},\ }\href {https://doi.org/10.1126/sciadv.aay8717} {\bibfield  {journal} {\bibinfo  {journal} {Science Advances}\ }\textbf {\bibinfo {volume} {6}},\ \bibinfo {pages} {eaay8717} (\bibinfo {year} {2020})}\BibitemShut {NoStop}%
\bibitem [{\citenamefont {Möller}\ \emph {et~al.}(2024)\citenamefont {Möller}, \citenamefont {Probst}, \citenamefont {Jansen}, \citenamefont {Schumacher}, \citenamefont {Brede}, \citenamefont {Dewhurst}, \citenamefont {Reutzel}, \citenamefont {Steil}, \citenamefont {Sharma},\ and\ \citenamefont {Mathias}}]{moller_verification_2024}%
  \BibitemOpen
  \bibfield  {author} {\bibinfo {author} {\bibfnamefont {C.}~\bibnamefont {Möller}}, \bibinfo {author} {\bibfnamefont {H.}~\bibnamefont {Probst}}, \bibinfo {author} {\bibfnamefont {G.~S.~M.}\ \bibnamefont {Jansen}}, \bibinfo {author} {\bibfnamefont {M.}~\bibnamefont {Schumacher}}, \bibinfo {author} {\bibfnamefont {M.}~\bibnamefont {Brede}}, \bibinfo {author} {\bibfnamefont {J.~K.}\ \bibnamefont {Dewhurst}}, \bibinfo {author} {\bibfnamefont {M.}~\bibnamefont {Reutzel}}, \bibinfo {author} {\bibfnamefont {D.}~\bibnamefont {Steil}}, \bibinfo {author} {\bibfnamefont {S.}~\bibnamefont {Sharma}},\ and\ \bibinfo {author} {\bibfnamefont {S.}~\bibnamefont {Mathias}},\ }\bibfield  {title} {\bibinfo {title} {Verification of ultrafast spin transfer effects in iron-nickel alloys},\ }\href {https://doi.org/10.1038/s42005-024-01555-3} {\bibfield  {journal} {\bibinfo  {journal} {Communications Physics}\ }\textbf {\bibinfo {volume} {7}},\ \bibinfo {pages} {1} (\bibinfo {year} {2024})}\BibitemShut {NoStop}%
\bibitem [{\citenamefont {von Korff~Schmising}\ \emph {et~al.}(2024)\citenamefont {von Korff~Schmising}, \citenamefont {Jana}, \citenamefont {Z\"ulich}, \citenamefont {Sommer},\ and\ \citenamefont {Eisebitt}}]{korffschmising_2024}%
  \BibitemOpen
  \bibfield  {author} {\bibinfo {author} {\bibfnamefont {C.}~\bibnamefont {von Korff~Schmising}}, \bibinfo {author} {\bibfnamefont {S.}~\bibnamefont {Jana}}, \bibinfo {author} {\bibfnamefont {O.}~\bibnamefont {Z\"ulich}}, \bibinfo {author} {\bibfnamefont {D.}~\bibnamefont {Sommer}},\ and\ \bibinfo {author} {\bibfnamefont {S.}~\bibnamefont {Eisebitt}},\ }\bibfield  {title} {\bibinfo {title} {Direct versus indirect excitation of ultrafast magnetization dynamics in {Fe}{Ni} alloys},\ }\href {https://doi.org/10.1103/PhysRevResearch.6.013270} {\bibfield  {journal} {\bibinfo  {journal} {Phys. Rev. Res.}\ }\textbf {\bibinfo {volume} {6}},\ \bibinfo {pages} {013270} (\bibinfo {year} {2024})}\BibitemShut {NoStop}%
\bibitem [{\citenamefont {Kuiper}\ \emph {et~al.}(2014)\citenamefont {Kuiper}, \citenamefont {Roth}, \citenamefont {Schellekens}, \citenamefont {Schmitt}, \citenamefont {Koopmans}, \citenamefont {Cinchetti},\ and\ \citenamefont {Aeschlimann}}]{kuiper_spin-orbit_2014}%
  \BibitemOpen
  \bibfield  {author} {\bibinfo {author} {\bibfnamefont {K.~C.}\ \bibnamefont {Kuiper}}, \bibinfo {author} {\bibfnamefont {T.}~\bibnamefont {Roth}}, \bibinfo {author} {\bibfnamefont {A.~J.}\ \bibnamefont {Schellekens}}, \bibinfo {author} {\bibfnamefont {O.}~\bibnamefont {Schmitt}}, \bibinfo {author} {\bibfnamefont {B.}~\bibnamefont {Koopmans}}, \bibinfo {author} {\bibfnamefont {M.}~\bibnamefont {Cinchetti}},\ and\ \bibinfo {author} {\bibfnamefont {M.}~\bibnamefont {Aeschlimann}},\ }\bibfield  {title} {\bibinfo {title} {Spin-orbit enhanced demagnetization rate in {Co}/{Pt}-multilayers},\ }\href {https://doi.org/10.1063/1.4902069} {\bibfield  {journal} {\bibinfo  {journal} {Applied Physics Letters}\ }\textbf {\bibinfo {volume} {105}},\ \bibinfo {pages} {202402} (\bibinfo {year} {2014})}\BibitemShut {NoStop}%
\bibitem [{\citenamefont {Hofherr}\ \emph {et~al.}(2018)\citenamefont {Hofherr}, \citenamefont {Moretti}, \citenamefont {Shim}, \citenamefont {Häuser}, \citenamefont {Safonova}, \citenamefont {Stiehl}, \citenamefont {Ali}, \citenamefont {Sakshath}, \citenamefont {Kim}, \citenamefont {Kim}, \citenamefont {Kim}, \citenamefont {Hong}, \citenamefont {Kapteyn}, \citenamefont {Murnane}, \citenamefont {Cinchetti}, \citenamefont {Steil}, \citenamefont {Mathias}, \citenamefont {Stadtmüller}, \citenamefont {Albrecht}, \citenamefont {Kim}, \citenamefont {Nowak},\ and\ \citenamefont {Aeschlimann}}]{hofherr_induced_2018}%
  \BibitemOpen
  \bibfield  {author} {\bibinfo {author} {\bibfnamefont {M.}~\bibnamefont {Hofherr}}, \bibinfo {author} {\bibfnamefont {S.}~\bibnamefont {Moretti}}, \bibinfo {author} {\bibfnamefont {J.}~\bibnamefont {Shim}}, \bibinfo {author} {\bibfnamefont {S.}~\bibnamefont {Häuser}}, \bibinfo {author} {\bibfnamefont {N.~Y.}\ \bibnamefont {Safonova}}, \bibinfo {author} {\bibfnamefont {M.}~\bibnamefont {Stiehl}}, \bibinfo {author} {\bibfnamefont {A.}~\bibnamefont {Ali}}, \bibinfo {author} {\bibfnamefont {S.}~\bibnamefont {Sakshath}}, \bibinfo {author} {\bibfnamefont {J.~W.}\ \bibnamefont {Kim}}, \bibinfo {author} {\bibfnamefont {D.~H.}\ \bibnamefont {Kim}}, \bibinfo {author} {\bibfnamefont {H.~J.}\ \bibnamefont {Kim}}, \bibinfo {author} {\bibfnamefont {J.~I.}\ \bibnamefont {Hong}}, \bibinfo {author} {\bibfnamefont {H.~C.}\ \bibnamefont {Kapteyn}}, \bibinfo {author} {\bibfnamefont {M.~M.}\ \bibnamefont {Murnane}}, \bibinfo {author} {\bibfnamefont {M.}~\bibnamefont {Cinchetti}}, \bibinfo {author} {\bibfnamefont
  {D.}~\bibnamefont {Steil}}, \bibinfo {author} {\bibfnamefont {S.}~\bibnamefont {Mathias}}, \bibinfo {author} {\bibfnamefont {B.}~\bibnamefont {Stadtmüller}}, \bibinfo {author} {\bibfnamefont {M.}~\bibnamefont {Albrecht}}, \bibinfo {author} {\bibfnamefont {D.~E.}\ \bibnamefont {Kim}}, \bibinfo {author} {\bibfnamefont {U.}~\bibnamefont {Nowak}},\ and\ \bibinfo {author} {\bibfnamefont {M.}~\bibnamefont {Aeschlimann}},\ }\bibfield  {title} {\bibinfo {title} {Induced versus intrinsic magnetic moments in ultrafast magnetization dynamics},\ }\href {https://doi.org/10.1103/PhysRevB.98.174419} {\bibfield  {journal} {\bibinfo  {journal} {Physical Review B}\ }\textbf {\bibinfo {volume} {98}},\ \bibinfo {pages} {174419} (\bibinfo {year} {2018})}\BibitemShut {NoStop}%
\bibitem [{\citenamefont {Yamamoto}\ \emph {et~al.}(2019)\citenamefont {Yamamoto}, \citenamefont {Kubota}, \citenamefont {Suzuki}, \citenamefont {Hirata}, \citenamefont {Carva}, \citenamefont {Berritta}, \citenamefont {Takubo}, \citenamefont {Uemura}, \citenamefont {Fukaya}, \citenamefont {Tanaka}, \citenamefont {Nishimura}, \citenamefont {Ohkochi}, \citenamefont {Katayama}, \citenamefont {Togashi}, \citenamefont {Tamasaku}, \citenamefont {Yabashi}, \citenamefont {Tanaka}, \citenamefont {Seki}, \citenamefont {Takanashi}, \citenamefont {Oppeneer},\ and\ \citenamefont {Wadati}}]{yamamoto_ultrafast_2019}%
  \BibitemOpen
  \bibfield  {author} {\bibinfo {author} {\bibfnamefont {K.}~\bibnamefont {Yamamoto}}, \bibinfo {author} {\bibfnamefont {Y.}~\bibnamefont {Kubota}}, \bibinfo {author} {\bibfnamefont {M.}~\bibnamefont {Suzuki}}, \bibinfo {author} {\bibfnamefont {Y.}~\bibnamefont {Hirata}}, \bibinfo {author} {\bibfnamefont {K.}~\bibnamefont {Carva}}, \bibinfo {author} {\bibfnamefont {M.}~\bibnamefont {Berritta}}, \bibinfo {author} {\bibfnamefont {K.}~\bibnamefont {Takubo}}, \bibinfo {author} {\bibfnamefont {Y.}~\bibnamefont {Uemura}}, \bibinfo {author} {\bibfnamefont {R.}~\bibnamefont {Fukaya}}, \bibinfo {author} {\bibfnamefont {K.}~\bibnamefont {Tanaka}}, \bibinfo {author} {\bibfnamefont {W.}~\bibnamefont {Nishimura}}, \bibinfo {author} {\bibfnamefont {T.}~\bibnamefont {Ohkochi}}, \bibinfo {author} {\bibfnamefont {T.}~\bibnamefont {Katayama}}, \bibinfo {author} {\bibfnamefont {T.}~\bibnamefont {Togashi}}, \bibinfo {author} {\bibfnamefont {K.}~\bibnamefont {Tamasaku}}, \bibinfo {author} {\bibfnamefont {M.}~\bibnamefont
  {Yabashi}}, \bibinfo {author} {\bibfnamefont {Y.}~\bibnamefont {Tanaka}}, \bibinfo {author} {\bibfnamefont {T.}~\bibnamefont {Seki}}, \bibinfo {author} {\bibfnamefont {K.}~\bibnamefont {Takanashi}}, \bibinfo {author} {\bibfnamefont {P.~M.}\ \bibnamefont {Oppeneer}},\ and\ \bibinfo {author} {\bibfnamefont {H.}~\bibnamefont {Wadati}},\ }\bibfield  {title} {\bibinfo {title} {Ultrafast demagnetization of {Pt} magnetic moment in {L}1$_{\textrm{0}}$ -{FePt} probed by magnetic circular dichroism at a hard x-ray free electron laser},\ }\href {https://doi.org/10.1088/1367-2630/ab5ac2} {\bibfield  {journal} {\bibinfo  {journal} {New Journal of Physics}\ }\textbf {\bibinfo {volume} {21}},\ \bibinfo {pages} {123010} (\bibinfo {year} {2019})}\BibitemShut {NoStop}%
\bibitem [{\citenamefont {Vaskivskyi}\ \emph {et~al.}(2021)\citenamefont {Vaskivskyi}, \citenamefont {Malik}, \citenamefont {Salemi}, \citenamefont {Turenne}, \citenamefont {Knut}, \citenamefont {Brock}, \citenamefont {Stefanuik}, \citenamefont {Söderström}, \citenamefont {Carva}, \citenamefont {Fullerton}, \citenamefont {Oppeneer}, \citenamefont {Karis},\ and\ \citenamefont {Dürr}}]{vaskivskyi_element-specific_2021}%
  \BibitemOpen
  \bibfield  {author} {\bibinfo {author} {\bibfnamefont {I.}~\bibnamefont {Vaskivskyi}}, \bibinfo {author} {\bibfnamefont {R.~S.}\ \bibnamefont {Malik}}, \bibinfo {author} {\bibfnamefont {L.}~\bibnamefont {Salemi}}, \bibinfo {author} {\bibfnamefont {D.}~\bibnamefont {Turenne}}, \bibinfo {author} {\bibfnamefont {R.}~\bibnamefont {Knut}}, \bibinfo {author} {\bibfnamefont {J.}~\bibnamefont {Brock}}, \bibinfo {author} {\bibfnamefont {R.}~\bibnamefont {Stefanuik}}, \bibinfo {author} {\bibfnamefont {J.}~\bibnamefont {Söderström}}, \bibinfo {author} {\bibfnamefont {K.}~\bibnamefont {Carva}}, \bibinfo {author} {\bibfnamefont {E.~E.}\ \bibnamefont {Fullerton}}, \bibinfo {author} {\bibfnamefont {P.~M.}\ \bibnamefont {Oppeneer}}, \bibinfo {author} {\bibfnamefont {O.}~\bibnamefont {Karis}},\ and\ \bibinfo {author} {\bibfnamefont {H.~A.}\ \bibnamefont {Dürr}},\ }\bibfield  {title} {\bibinfo {title} {Element-{Specific} {Magnetization} {Dynamics} in {Co}–{Pt} {Alloys} {Induced} by {Strong} {Optical} {Excitation}},\
  }\href {https://doi.org/10.1021/acs.jpcc.1c02311} {\bibfield  {journal} {\bibinfo  {journal} {The Journal of Physical Chemistry C}\ }\textbf {\bibinfo {volume} {125}},\ \bibinfo {pages} {11714} (\bibinfo {year} {2021})}\BibitemShut {NoStop}%
\bibitem [{\citenamefont {Hennes}\ \emph {et~al.}(2022)\citenamefont {Hennes}, \citenamefont {Lambert}, \citenamefont {Chardonnet}, \citenamefont {Delaunay}, \citenamefont {Chiuzbăian}, \citenamefont {Jal},\ and\ \citenamefont {Vodungbo}}]{hennes_element-selective_2022}%
  \BibitemOpen
  \bibfield  {author} {\bibinfo {author} {\bibfnamefont {M.}~\bibnamefont {Hennes}}, \bibinfo {author} {\bibfnamefont {G.}~\bibnamefont {Lambert}}, \bibinfo {author} {\bibfnamefont {V.}~\bibnamefont {Chardonnet}}, \bibinfo {author} {\bibfnamefont {R.}~\bibnamefont {Delaunay}}, \bibinfo {author} {\bibfnamefont {G.~S.}\ \bibnamefont {Chiuzbăian}}, \bibinfo {author} {\bibfnamefont {E.}~\bibnamefont {Jal}},\ and\ \bibinfo {author} {\bibfnamefont {B.}~\bibnamefont {Vodungbo}},\ }\bibfield  {title} {\bibinfo {title} {Element-selective analysis of ultrafast demagnetization in {Co}/{Pt} multilayers exhibiting large perpendicular magnetic anisotropy},\ }\href {https://doi.org/10.1063/5.0080275} {\bibfield  {journal} {\bibinfo  {journal} {Applied Physics Letters}\ }\textbf {\bibinfo {volume} {120}},\ \bibinfo {pages} {072408} (\bibinfo {year} {2022})}\BibitemShut {NoStop}%
\bibitem [{dat()}]{dataJR}%
  \BibitemOpen
  \bibfield  {journal} {\bibinfo  {journal} {zenodo.org}\ }\href {https://doi.org/10.5281/zenodo.15463833} {10.5281/zenodo.15463833}\BibitemShut {NoStop}%
\end{thebibliography}%

\end{document}